\documentclass[journal,twocolumn]{IEEEtran}
\IEEEoverridecommandlockouts
\usepackage{amsmath,amssymb,amsfonts}
\usepackage{algorithmic}
\usepackage{amsmath} 
\usepackage{graphicx}
\usepackage{textcomp}
\usepackage{xcolor}
\usepackage{amsmath}
\renewcommand{\vec}[1]{\mathbf{#1}}
\usepackage{algorithm}
\usepackage{algorithmic}
\usepackage{graphicx} 
\usepackage{float} 
\usepackage{stfloats}
\usepackage[section]{placeins}
\usepackage{subfigure}
\usepackage{booktabs}
\usepackage[numbers,sort&compress]{natbib}
\usepackage{amssymb}
\usepackage{setspace}
\usepackage{multirow}
\usepackage{color}
\usepackage{hyperref}
\usepackage{bm}
\fontsize{5.0pt}{\baselineskip}\selectfont
\usepackage{epsfig}
\def\BibTeX{{\rm B\kern-.05em{\sc i\kern-.025em b}\kern-.08em
    T\kern-.1667em\lower.7ex\hbox{E}\kern-.125emX}}

\begin{document}
\addtolength{\topmargin}{0.04in}
\title{ Mixed-ADC Based PMCW  MIMO Radar Angle-Doppler Imaging}

\author{Xiaolei~Shang, Ronghao~Lin and  Yuanbo~Cheng 
\thanks{This work was supported in part by the Youth Innovation Promotion Association CAS (CX2100060053), in part by the Anhui Provincial Natural Science Foundation under Grant 2208085J17, and in part by the National Natural Science Foundation of China under Grant 62271461.  
\par X. Shang, R. Lin and Y. Cheng are  with the Department of Electronic Engineering and Information
Science, University of Science and Technology of China, Hefei 230027, China
(e-mail: xlshang@mail.ustc.edu.cn, lrh23703@mail.ustc.edu.cn and cyb967@mail.ustc.edu.cn).}}

\maketitle
\begin{abstract}
Phase-modulated continuous-wave (PMCW)  multiple-input multiple-output (MIMO) radar systems are known to possess excellent mutual interference mitigation capabilities, but require costly and power-hungry high sampling rate and high-precision analog-to-digital converters  (ADC's).  To reduce cost and power consumption, we consider a  mixed-ADC architecture, in which most receive antenna outputs are sampled by one-bit ADC's, and only one  or a few outputs by high-precision ADC's.  We first derive the Cram{\'e}r-Rao bound (CRB)  for the  mixed-ADC based PMCW MIMO radar to characterize the  best achievable performance of an unbiased target parameter estimator. The   CRB analysis demonstrates that  the mixed-ADC architecture with  a relatively small number of high-precision ADC's and a large number of one-bit ADC's  allows us to drastically reduce the hardware cost and power consumption while still maintain a high dynamic range needed for autonomous driving applications.   We also introduce a two-step estimator to realize the computationally efficient maximum likelihood (ML) estimation of the target parameters. We  formulate the angle-Doppler imaging problem   as a sparse parameter estimation problem, and a computationally efficient majorization-minimization (MM) based  estimator of sparse parameters,  referred to as mLIKES, is devised for accurate angle-Doppler imaging. This is followed by using a relaxation-based approach to cyclically refine the results of mLIKES for accurate off-grid target  parameter estimation.  Numerical examples  are provided to demonstrate the effectiveness of  the proposed algorithms for angle-Doppler imaging using mixed-ADC based  PMCW MIMO radar.
\end{abstract}

\section{Introduction}
Millimeter-wave radar is an indispensable sensor for automotive radar  applications, including advanced driver assistance and  autonomous driving. Compared with other sensors, such as camera and lidar, radar can provide excellent sensing capabilities even in poor lighting or adverse weather conditions \cite{sun2020mimo}. 
\par Multiple-input multiple-output (MIMO) radar  can achieve high angular resolution with a relatively  small number of antennas \cite{bliss2003multiple,li2007mimo,li2009mimo}, and  MIMO radar has become a standard  for  automotive radar applications. Because of the low-cost advantage, most of the existing automotive radar  are linear frequency-modulated continuous-wave (LFMCW) MIMO radar systems. However, with the increasing popularity of automotive radar, the mutual interference problems become increasingly severe \cite{alland2019interference}.  Due to the code-division multiple access (CDMA) type of interference mitigation advantages, phase-modulated continuous-wave (PMCW) MIMO radar systems are attracting increasing attention in the literature \cite{alland2019interference,sun2020mimo}.  However, compared with LFMCW systems, PMCW systems require the receiver analog-to-digital converters (ADC's) to have a much higher sampling rate \cite{sun2020mimo}.
\par PMCW MIMO radar systems are typically required to sample the received signal at twice the bandwidth of the transmitted signal. Automotive radar systems typically operate in the millimeter-wave frequency range between 77-81 GHz, with a maximum bandwidth of 4 GHz \cite{alland2019interference}. At higher carrier frequencies, such as at 140 GHz under development by IMEC \cite{banerjee2019millimeter}, the radar bandwidth can be much larger than the current maximum of 4 GHz. However, high rate and high precision ADC's are both expensive and power hungry \cite{walden1999analog,cheng2022interval}.  High level autonomous driving requires high angular resolution, making it necessary to deploy a large number of receive antennas, and hence a large  number of ADC's, further increasing the cost and power consumption of the PMCW MIMO automotive radar. 
\par Using low quantization resolution ADC's (e.g., 1-4 bits) at the receiver \cite{bar2002doa,fang2014sparse,zhao2019one,shang2021weighted,wu2022parameter}  is a promising approach to mitigate the aforementioned ADC problems because of their low cost and low power consumption advantages. However, low-precision ADC's suffer from dynamic range problems, i.e., a strong target can mask a weak target  due to the high quantization errors \cite{walden1999analog,stoica2021cramer}.  Yet  high dynamic range is required for  high-performance automotive radar applications.  To mitigate the dynamic range problems of low-bit ADC's, we consider herein a  mixed-ADC based architecture, in which most receive antenna outputs are sampled by one-bit ADC's, and one or a few outputs by high-resolution ADC's, for PMCW MIMO radar systems. The  main contributions of this paper can be summarized as follows:

We first derive the  Cram{\'e}r-Rao bound (CRB)  for the mixed-ADC based architecture for PMCW MIMO radar to characterize its best achievable performance for an unbiased target parameter estimator. Moreover, upper and lower bounds are provided on the CRB of  the mixed-ADC system. The  CRB analysis demonstrates that  the mixed-ADC based architecture with a relatively small number  of high-precision ADC's and a large number of one-bit ADC's can  overcome  the dynamic range problems of one-bit or low-bit sampling.

We next introduce a two-step estimator to realize the computationally efficient maximum likelihood (ML) estimation of the target parameters. Firstly, we formulate the angle-Doppler estimation problem encountered in the mixed-ADC based system as a sparse parameter estimation problem, and introduce a  likelihood-based estimation approach,  referred to as the mLIKES algorithm, for accurate angle-Doppler imaging. { We use the
majorization-minimization (MM) technique \cite{hunter2004tutorial,stoica2004cyclic,mairal2015incremental,hong2015unified,sun2016majorization},  which can transform a difficult optimization problem into  a  sequence  of much simpler ones, to simply the optimization problem encountered in mLIKES to attain enhanced computational efficiency.} By making use of the majorization-minimization technique \cite{sun2016majorization,stoica2004cyclic}, the parameters can be updated iteratively. Moreover, the Nesterov acceleration technique \cite{nesterov1983method,beck2009fast} is adopted to  accelerate the convergence rate of the MM iterations. Furthermore, a RELAX-based approach is devised to further refine the results of mLIKES. Finally, we use  numerical examples to demonstrate that   mLIKES  and RELAX-based refinement can be combined to obtain  accurate target parameter estimates, with accuracy  approaching the corresponding CRB.

The rest of this paper is organized as follows. A mixed-ADC based PMCW MIMO radar system is introduced in Section II. The CRB for the target parameters is derived in Section III. We introduce a two-step estimator  to accurate  target angle-Doppler parameter estimation in Section IV. The numerical examples are presented in Section V and  Section VI concludes this paper.

\indent \textit{Notation:} We denote vectors and matrices by bold lowercase and uppercase letters, respectively. $(\cdot)^{*}$, $(\cdot)^T$ and $(\cdot)^H$ represent the complex conjugate, transpose and the conjugate transpose, respectively. $\vec{R}\in \mathbb{R}^{N\times M}$ or  $\vec{R}\in \mathbb{C}^{N\times M}$ denotes a real or complex-valued $N\times M$ matrix $\vec{R}$. $\vec{A}\otimes\vec{B}$, $\vec{A}\odot\vec{B}$ and $\vec{A} \oplus \vec{B}$ denote, respectively,  the Kronecker, Hadamard and Khatri–Rao matrix products. $\vec{A}\left[\bm{\delta},:\right]$ and  $\vec{a}(\bm{\delta})$, respectively,  extract rows and elements  corresponding to the non-zero values of the logical array $\bm{\delta}$.
${\rm vec}(\cdot)$ refers to the column-wise  vectorization operation and ${\rm diag}(\vec{d})$ denotes a diagonal matrix with diagonal entries formed  from $\vec{d}$. 
${\rm tr}\lbrace \vec{R}\rbrace$ and $|\vec{R}|$, respectively, denote the trace and determinant of a square matrix $\vec{R}$.  $\vec{A} \succeq \vec{B}$ means that $\vec{A}-\vec{B}$ is a positive semidefinite matrix.
Denote $\vec{1}_N=\begin{bmatrix}1,\dots,1 \end{bmatrix}^T \in \mathbb{R}^{N\times 1}$. $\vec{I}_N$  denotes the $N \times N$ identity matrix and $\vec{e}_n$ is the $n$-th column of $\vec{I}_N$. $A_{\rm R}=\Re \lbrace {A} \rbrace$ and ${A}_{\rm I}=\Im \lbrace A \rbrace$, where $\Re \lbrace \cdot \rbrace$ and $\Im \lbrace \cdot \rbrace$ denote taking the real  and imaginary parts, respectively. Finally, $j\triangleq \sqrt{-1}$.

\section{Problem formulation}
\begin{figure}[htb]
    \centering
    \includegraphics[width=0.43\textwidth]{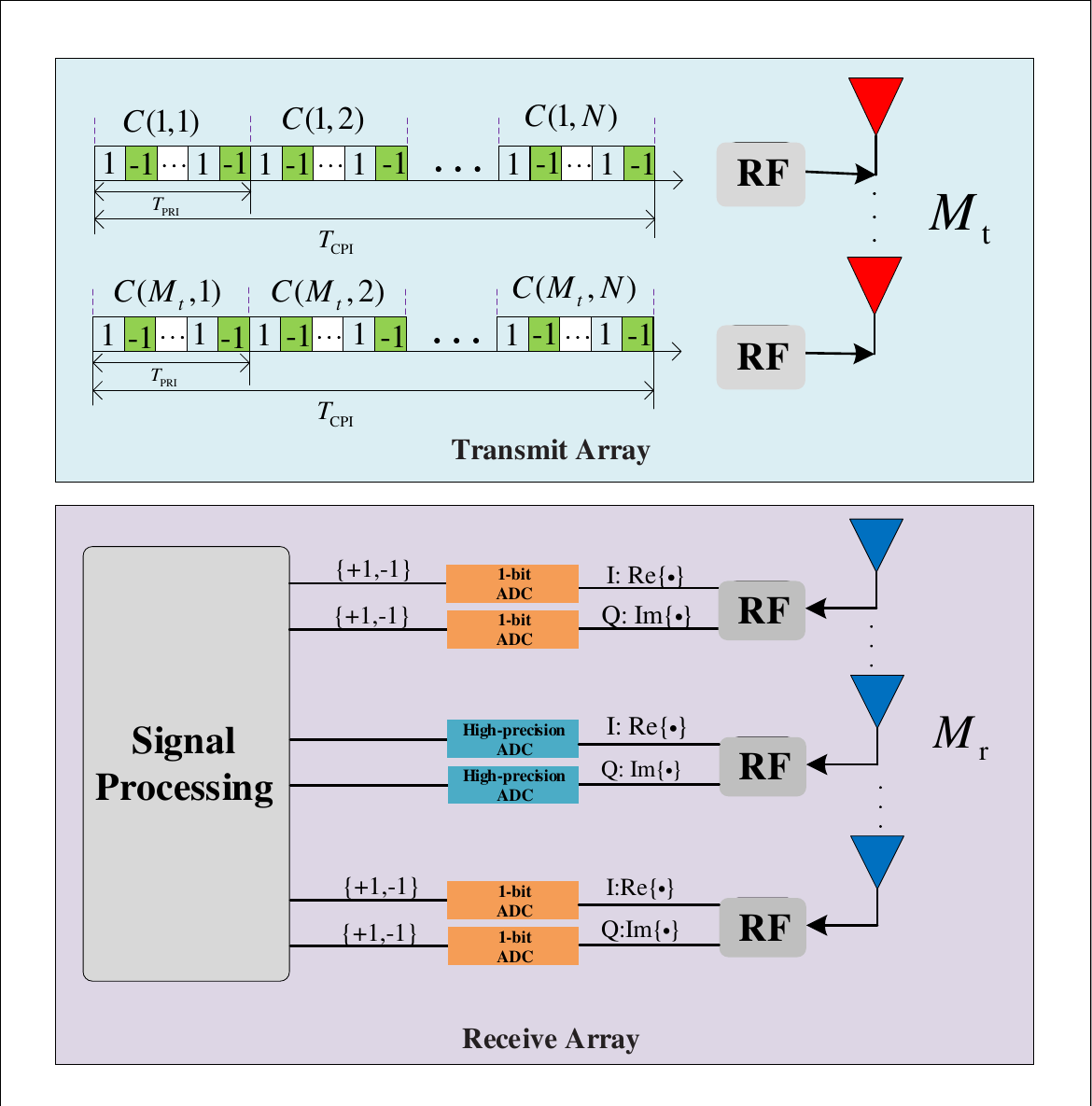}
    \caption{The mixed-ADC based PMCW MIMO radar architecture.}
    \label{fig:sysmodel}
\end{figure}

Consider a  PMCW MIMO radar system with $M_{\rm t}$ transmit antennas and $M_{\rm r}$ receive antennas. As shown in Fig. \ref{fig:sysmodel}, the PMCW MIMO radar uses $M_{\rm t}$ antennas to simultaneously transmit the same fast-time binary sequence after multiplying it with a unimodular random slow-time code  that is unique for each transmit antenna and  changes from one pulse repetition interval (PRI) to another within the coherent processing interval (CPI) \footnote{Herein waveform orthogonality
is achieved by using slow-time code division multiplexing, which  uses low-rate pseudorandom codes, changing  from one
PRI to another. This approach is low-cost and may be preferred in diverse applications.}. Let $\vec{a}_{\rm t}(\theta)$ and $\vec{a}_{\rm r}(\theta)$, respectively, denote the transmit and receive steering vectors for a target at an azimuth angle $\theta$:
\begin{equation}
    \vec{a}_{\rm t}(\theta)=\begin{bmatrix} 1, e^{-j2\pi\frac{d_{\rm t}}{\lambda}\sin\theta},  \dots, e^{-j2\pi\frac{d_{\rm t}}{\lambda}(M_{\rm t}-1)\sin\theta}\end{bmatrix}^T \in \mathbb{C}^{M_{\rm t}\times 1}
\end{equation}
and
\begin{equation}
    \vec{a}_{\rm r}(\theta)=\begin{bmatrix} 1, e^{-j2\pi\frac{d_{\rm r}}{\lambda}\sin\theta},  \dots, e^{-j2\pi\frac{d_{\rm r}}{\lambda}(M_{\rm r}-1)\sin\theta}\end{bmatrix}^T \in \mathbb{C}^{M_{\rm r }\times 1}
\end{equation}
where $\lambda$ is the wavelength; $d_{\rm t}$ and $d_{\rm r}$ represent the inter-element distances of the transmit and receive uniform linear  arrays, respectively. Assume that a total number of  $N$ PRI's are used during a CPI to determine the Doppler shifts of moving targets. A nominal slow-time  temporal steering vector $\vec{d}({\omega})$ for a target with Doppler frequency shift $\omega$ can be written as:
\begin{equation}
    \vec{d}({\omega})=\begin{bmatrix} 1, e^{j\omega},  \dots, e^{j(N-1)\omega} \end{bmatrix}^T \in \mathbb{C}^{N\times 1}.
\end{equation}
At  a given fixed fast-time, the data received by such a PMCW MIMO radar in the slow-time and angular domain can be written as:
\begin{equation}
    \vec{X}=\sum_{k=1}^K \vec{a}_{\rm r}(\theta_k)b_k (\vec{C}^T\vec{a}_{\rm t}(\theta_k) \odot \vec{d}(\omega_k))^T +\vec{E} \in \mathbb{C}^{M_{\rm r} \times N}, \label{eq:echo}
\end{equation}
where $\lbrace \theta_k \rbrace_{k=1}^K$ and $\lbrace \omega_k \rbrace_{k=1}^K$ represent the azimuth angles and Doppler shifts  of $K$ targets, respectively; $\lbrace b_k \rbrace_{k=1}^K$ are the  complex-valued  amplitudes, which  are proportional to the radar-cross-sections (RCS) of the targets; $\vec{C} \in \mathbb{C}^{M_{\rm t} \times N}$ represents the slow-time code matrix, with the  $(m,n)$th element, $C(m,n)=e^{j\phi_{m,n}}$, denoting the slow-time code for the $m$th antenna at the $n$th PRI;  and  $\vec{E}$ is the unknown additive noise  matrix. The extension to the case of joint range-Doppler-angle estimation is not difficult but it leads to more complicated expressions. To keep the exposition herein as simple as possible, we will only consider angle-Doppler imaging herein. 

The observed data matrix can also be expressed compactly as follows:
\begin{align}
     \vec{X}&=\vec{A}_{\rm r}\vec{B}\left( \vec{C}^T\vec{A}_{\rm t}\odot \vec{D}\right)^T + \vec{E}  \nonumber \\
    &=\vec{A}_{\rm r}\vec{B}\vec{V}^T+\vec{E}, \label{eq:model}
\end{align}
where 
\begin{align}
\vec{A}_{\rm r}&=\begin{bmatrix}\vec{a}_{\rm r}(\theta_1),\dots, \vec{a}_{\rm r}(\theta_K) \end{bmatrix} \in \mathbb{C}^{M_{\rm r} \times K}, \nonumber   \\
\vec{A}_{\rm t}&=\begin{bmatrix}\vec{a}_{\rm t}(\theta_1),\dots, \vec{a}_{\rm t}(\theta_K) \end{bmatrix} \in \mathbb{C}^{M_{\rm t} \times K}, \nonumber  \\
  \vec{D}&=\begin{bmatrix}\vec{d}(\omega_1), \dots, \vec{d}(\omega_K) \end{bmatrix} \in \mathbb{C}^{N \times K}, \nonumber \\
  \vec{V}&\triangleq \vec{C}^T\vec{A}_{\rm t} \odot \vec{D} \in \mathbb{C}^{N \times K}, \nonumber \\
\end{align}
and
\begin{align}
\vec{b}&=\begin{bmatrix}b_1,\dots, b_K \end{bmatrix}^T, \nonumber \\
\vec{B}&={\rm diag}(\vec{b}).
\end{align}
Let the vectors $\bm{\theta}$ and $\bm{\omega}$ collect the target azimuth angles and Doppler frequencies, respectively. When there is no risk for confusion, we  omit the dependence of different functions (such as $\vec{A}_{\rm r}(\bm{\theta})$ on $\bm{\theta}$)  to simplify the notation.
Let $\vec{x}={\rm vec}(\vec{X})$ and $\vec{e}={\rm vec}(\vec{E})$. The data model in \eqref{eq:model}  is equivalent to:
\small
\begin{align}
    \vec{x}&=\sum_{k=1}^K b_k \vec{v}(\theta_k,\omega_k) \otimes \vec{a}_{\rm r}(\theta_k) + \vec{e} \nonumber \\
    &= \sum_{k=1}^K b_k \left[\left( \vec{C}^T\vec{a}_{\rm t}(\theta_k) \right)\odot \vec{d}(\omega_k) \right] \otimes \left[ \vec{a}_{\rm r}(\theta_k) \odot \vec{1}_{M_{\rm r}} \right] + \vec{e} \nonumber \\
    &= \sum_{k=1}^K b_k \left[ \left( \vec{C}^T \vec{a}_{\rm t}(\theta_k) \right)\otimes \vec{a}_{\rm r}(\theta_k) \right]\odot \left[\vec{d}(\omega_k) \otimes \vec{1}_{M_{\rm r}} \right] + \vec{e} \nonumber \\
    &=\sum_{k=1}^K b_k \left[ \left( \vec{C}^T \otimes \vec{I}_{M_{\rm r}}\right) \left( \vec{a}_{\rm t}(\theta_k)\otimes \vec{a}_{\rm r}(\theta_k) \right)\right] \odot \left[\vec{d}(\omega_k) \otimes \vec{1}_{M_{\rm r}} \right] +\vec{e}, \label{eq:vecmodel}
\end{align}
\normalsize
where $\vec{v}(\theta_k,\omega_k)$ is the $k$th column of $\vec{V}$; and the third and fourth equalities above, respectively,  are obtained by using  the following mixed-product properties: 
\begin{align}
    \left(\vec{A}\odot\vec{C}\right)\otimes \left(\vec{B} \odot \vec{D} \right)&=\left( \vec{A} \otimes \vec{B} \right)\odot \left( \vec{C} \otimes \vec{D}\right), 
\end{align}
and
\begin{align}
    \left(\vec{A}\vec{C}\right)\otimes \left(\vec{B}\vec{D}\right)&=\left(\vec{A}\otimes \vec{B} \right)\left( \vec{C}\otimes \vec{D}\right). \label{eq:mixedprod2}
\end{align}
Note from \eqref{eq:vecmodel} that the virtual aperture $\vec{a}_{\rm t}(\theta_k)\otimes \vec{a}_{\rm r}(\theta_k)$ is linearly mapped to $\left(\vec{C}^T \otimes \vec{I}_{M_{\rm r}} \right)\left(\vec{a}_{\rm t}(\theta_k) \otimes \vec{a}_{\rm r}(\theta_k) \right)$ by the matrix $\vec{C}^T\otimes \vec{I}_{M_{\rm r}}$, but the angle and Doppler information are coupled via the Hadamard product.

\subsection{One-bit Quantization}
When  all antennas adopt one-bit quantization at the receivers, we obtain the signed measurement matrix $\vec{Z}=\vec{Z}_{\rm R}+j\vec{Z}_{\rm I}$ by comparing the unquantized
received signal $\vec{X}$ with a known time-varying  threshold $\vec{H}\in \mathbb{C}^{M_{\rm r}\times N}$ \cite{shang2021weighted}:
\begin{equation}
    \vec{Z}={\rm signc}(\vec{X}-\vec{H}),
\end{equation}
where ${\rm signc}(x)={\rm sign}\left({\rm Re}[x]\right)+j{\rm sign}\left({\rm Im}[x]\right)$  and ${\rm sign}(x)$ is the element-wise sign operator defined as:
\begin{align}
{\rm sign}(x)=\begin{cases} \label{eq:window}
    \ 1 & x \geqslant 0, \\
    -1 & x<0.
    \end{cases}
\end{align}
\subsection{Mixed-ADC Quantization}
As shown in Fig. \ref{fig:sysmodel}, we also consider a mixed-ADC based  architecture in which only $L$ pairs of  ADC's are high-precision and all  other $(M_{\rm r}-L)$  pairs are one-bit ADC's. For the convenience of subsequent processing,  we define a high-precision ADC indicator vector $\bm{\delta}=\begin{bmatrix}\delta_1, \dots,  \delta_{M_{\rm r}} \end{bmatrix}^T$ with $\delta_m\in \lbrace 0,1\rbrace$, where  $\delta_m = 1$ means that the $m$-th antenna output is sampled by a pair of high-precision ADC's (for its I (in-phase) and Q (quadrature) channels), whereas $\delta_m= 0$  indicates one-bit ADC's. Then, the mixed-ADC quantized output is:
\begin{equation}
    \vec{Y}={\rm diag}(\bm{\delta})\vec{X} + {\rm diag}(\bar{\bm{\delta}})\vec{Z}, 
\end{equation}
where $\bar{\bm{\delta}}=\vec{1}_{M_{\rm r}}-\bm{\delta}$ is a one-bit ADC indicator.  For later use, we separate mixed-ADC outputs into the  high-precision  group $\vec{Y}_0=\vec{Y}\left[\bm{\delta},: \right]\in \mathbb{C}^{L\times N}$ and the one-bit group $\vec{Y}_1=\vec{Y}\left[\bar{\bm{\delta}},: \right]\in \mathbb{C}^{(M_{\rm r}-L)\times N}$. Let  $\vec{y}_0={\rm vec}(\vec{Y}_0)$ and $\vec{y}_1={\rm vec}(\vec{Y}_1)$. Then $\vec{y}_0$ and $\vec{y}_1$, respectively, have the following expressions:
\begin{align}
    \vec{y}_0&=\sum_{k=1}^K b_k \vec{a}_0(\theta_k,\omega_k) + \vec{e}_0 \in \mathbb{C}^{NL\times 1}, \nonumber
\end{align}
and
\begin{align}
        \vec{y}_1&={\rm signc}(\sum_{k=1}^K b_k\vec{a}_1(\theta_k,\omega_k) -\vec{h}_1 + \vec{e}_1) \in \mathbb{C}^{N(M_r-L)\times 1}, \label{eq:vecmodelnew}
\end{align}
where $\vec{a}_0(\theta_k,\omega_k)\triangleq \vec{v}(\theta_k,\omega_k)\otimes \tilde{\vec{a}}_{{\rm r}}(\theta_k)$; $\vec{a}_1(\theta_k,\omega_k)\triangleq \vec{v}(\theta_k,\omega_k)\otimes \check{\vec{a}}_{{\rm r}}(\theta_k)$;   $\tilde{\vec{a}}_{\rm r}(\theta_k)$ and $\check{\vec{a}}_{\rm r}(\theta_k)$ are  sub-vectors extracted from $\vec{a}_{\rm r}(\theta_k)$ via the vectors $\bm{\delta}$ and $\bar{\bm{\delta}}$, respectively; $\vec{e}_0$ and $\vec{e}_1$ represent the additive noise;  and $\vec{h}_1\triangleq {\rm vec}(\vec{H}\left[\bar{\bm{\delta}},:\right])$ is the time-varying threshold vector adopted by one-bit ADC's. 
\par  Our problem of interest herein is to estimate the unknown target parameters and form angle-Doppler images from  the mixed-ADC output matrix $\vec{Y}$. 
\section{Cram{\'e}r-Rao Bound}
Let $\bm{\varphi}$ collect all the real-valued unknown target parameters, i.e., $\bm{\varphi}=\begin{bmatrix} \bm{\theta}^T, \ \bm{\omega}^T,\ \vec{b}_{\rm R}^T, \ \vec{b}_{\rm I}^T \end{bmatrix}^T\in \mathbb{R}^{4K\times 1}$. When the noise power $\sigma^2$ is unknown,  the unknown parameter vector becomes $\bm{\zeta}=\begin{bmatrix}\bm{\varphi}^T, \ \sigma\end{bmatrix}^T\in \mathbb{R}^{(4K+1)\times 1}$. Under the assumption that $\vec{E}$ is the circularly symmetric complex-valued white Gaussian noise with  i.i.d. $\mathcal{CN}(0,\sigma^2)$ entries, the log-likelihood function of the measurement matrix $\vec{Y}$ is given by:
\begin{align}
    \ln L(\bm{\zeta})=\ln L_{0}(\bm{\zeta}) + \ln L_{1}(\bm{\zeta}), \label{eq:mixedml}
\end{align}
where $L_{0}(\bm{\zeta})$ and $L_{1}(\bm{\zeta})$ represent the likelihood functions of the high-precision and one-bit measurements, respectively.  The Fisher information matrix (FIM) for the mixed-ADC system can be written as:
\begin{equation}
    \vec{F}_{\rm m}={\rm E}\left[\frac{\partial\ln L(\bm{\zeta})}{\partial \bm{\zeta}} \frac{\partial \ln L(\bm{\zeta})}{\partial \bm{\zeta}^T} \right].
\end{equation}
It follows from \eqref{eq:mixedml} that 
\begin{small}
\begin{align}
    \vec{F}_{\rm m}=&{\rm E}\left[ \frac{\partial \ln L_{0}(\bm{\zeta})}{\partial \bm{\zeta}}\frac{\partial \ln L_{0}(\bm{\zeta})}{\partial \bm{\zeta}^T}\right] + {\rm E}\left[ \frac{\partial \ln L_{1}(\bm{\zeta})}{\partial \bm{\zeta}}\frac{\partial \ln L_{1}(\bm{\zeta})}{\partial \bm{\zeta}^T}\right] \nonumber  \\
    &+  {\rm E}\left[ \frac{\partial \ln L_{0}(\bm{\zeta})}{\partial \bm{\zeta}}\frac{\partial \ln L_{1}(\bm{\zeta})}{\partial \bm{\zeta}^T}\right]+  {\rm E}\left[ \frac{\partial \ln L_{1}(\bm{\zeta})}{\partial \bm{\zeta}}\frac{\partial \ln L_{0}(\bm{\zeta})}{\partial \bm{\zeta}^T}\right]. \label{eq:fm}
\end{align}
\end{small}
Using the fact that the noise entries $\lbrace E(m,n) \rbrace$ are independent random variables, we obtain:
\begin{align}
     {\rm E}\left[ \frac{\partial \ln L_{0}(\bm{\zeta})}{\partial \bm{\zeta}}\frac{\partial \ln L_{1}(\bm{\zeta})}{\partial \bm{\zeta}^T}\right]&=\underbrace{{\rm E}\left[ \frac{\partial \ln L_{0}(\bm{\zeta})}{\partial \bm{\zeta}}\right]}_{=\vec{0}} \underbrace{{\rm E}\left[\frac{\partial \ln L_{1}(\bm{\zeta})}{\partial \bm{\zeta}^T} \right]}_{=\vec{0}} \nonumber \\
     &=\vec{0}. \label{eq:cross}
\end{align}
Inserting \eqref{eq:cross} into \eqref{eq:fm} yields the following expression for $\vec{F}_{\rm m}$:
\begin{equation}
    \vec{F}_{\rm m}={\rm E}\left[ \frac{\partial \ln L_{0}(\bm{\zeta})}{\partial \bm{\zeta}}\frac{\partial \ln L_{0}(\bm{\zeta})}{\partial \bm{\zeta}^T}\right] + {\rm E}\left[ \frac{\partial \ln L_{1}(\bm{\zeta})}{\partial \bm{\zeta}}\frac{\partial \ln L_{1}(\bm{\zeta})}{\partial \bm{\zeta}^T}\right]. \label{eq:Fm}
\end{equation}
Equation \eqref{eq:Fm} demonstrates the fact that the FIM for the mixed-ADC outputs is the summation of the FIMs for the high-precision outputs and one-bit outputs. In the next two subsections, we focus on the derivations of the FIMs for high-precision outputs and one-bit outputs. Then, the FIM for the mixed-ADC outputs can be readily obtained.  
\vspace{-0.3cm}
\subsection{Cram{\'e}r-Rao Bound for High-precision Data}
\subsubsection{Known noise variance} The FIM for the high-precision outputs with respect to $\bm{\varphi}$ in \eqref{eq:echo} can be written as (see Appendix A for the detailed derivations):
\begin{equation}
    \vec{F}_{0}(\bm{\varphi})=\frac{2}{\sigma^2} \Re \left\{ \left[
 \begin{matrix}
    \vec{F}_{11} & \vec{F}_{12} & \vec{F}_{13} & j\vec{F}_{13}  \\
    \vec{F}_{12}^T & \vec{F}_{22} & \vec{F}_{23} & j\vec{F}_{23} \\
    \vec{F}_{13}^T & \vec{F}_{23}^T & \vec{F}_{33} & j\vec{F}_{33} \\
   j\vec{F}_{13}^T & j\vec{F}_{23}^T & j\vec{F}_{33}^T & \vec{F}_{33} \\
  \end{matrix} 
\right] \right\}, \label{eq:F0}
\end{equation} 
where 
\begin{footnotesize}
\begin{align}
    \vec{F}_{11}&=(\dot{\vec{A}}_{\rm r}^H\dot{\vec{A}}_{\rm r})\odot (\vec{B}^{*}\vec{V}^H\vec{V}\vec{B}) +(\dot{\vec{A}}_{\rm r}^H\vec{A}_{\rm r})\odot(\vec{B}^{*}\vec{V}^H\dot{\vec{V}}_{\bm{\theta}}\vec{B}) \nonumber \\
    & \quad + ({\vec{A}_{\rm r}^H\dot{\vec{A}}_{\rm r}}) \odot (\vec{B}^*\dot{\vec{V}}_{\bm{\theta}}^H\vec{V}\vec{B})+ (\vec{A}_{\rm r}^H\vec{A}_{\rm r})\odot(\vec{B}^*\dot{\vec{V}}^H_{\bm{\theta}}\dot{\vec{V}}_{\bm{\theta}}\vec{B}), \label{eq:F11} \\
    \vec{F}_{12}&=(\dot{\vec{A}}_{\rm r}^H\vec{A}_{\rm r})\odot (\vec{B}^*\vec{V}^H\dot{\vec{V}}_{\bm{\omega}}\vec{B}) + (\vec{A}_{\rm r}^H\vec{A}_{\rm r}) \odot (\vec{B}^*\dot{\vec{V}}_{\bm{\theta}}^H\dot{\vec{V}}_{\bm{\omega}}\vec{B}), \label{eq:F12} \\
    \vec{F}_{13}&=(\dot{\vec{A}}_{\rm r}^H\vec{A}_{\rm r})\odot (\vec{B}^*\vec{V}^H\vec{V})+ (\vec{A}_{\rm r}^H\vec{A}_{\rm r}) \odot (\vec{B}^H\dot{\vec{V}}_{\bm{\theta}}\vec{V}), \\
    \vec{F}_{22}&=(\vec{A}_{\rm r}^H\vec{A}_{\rm r}) \odot (\vec{B}^*\dot{\vec{V}}_{\bm{\omega}}^H \dot{\vec{V}}_{\bm{\omega}}\vec{B}), \\
    \vec{F}_{23}&=(\vec{A}_{\rm r}^H\vec{A}_{\rm r}) \odot (\vec{B}^*\dot{\vec{V}}^H_{\bm{\omega}}\vec{V}), \\
    \vec{F}_{33}&=(\vec{A}_{\rm r}^H\vec{A}_{\rm r}) \odot (\vec{V}^H\vec{V}), \label{eq:F33}
\end{align} 
\end{footnotesize}
and
\begin{align}
    \dot{\vec{A}}_{\rm r}&=\begin{bmatrix} \frac{\partial \vec{a}_{\rm r}(\theta_1)}{\partial \theta_1}, \dots,  \frac{ \partial \vec{a}_{\rm r}(\theta_K)}{\partial \theta_K} \end{bmatrix}, \label{eq:dAr} \\ 
    \dot{\vec{V}}_{\bm{\theta}}&=\begin{bmatrix} \frac{\partial \vec{v}(\theta_1,\omega_1)}{\partial \theta_1}, \dots,  \frac{ \partial \vec{v}(\theta_K,\omega_K)}{\partial \theta_K}  
    \end{bmatrix}, \\
    \dot{\vec{V}}_{\bm{\omega}}&=\begin{bmatrix} \frac{\partial \vec{v}(\theta_1,\omega_1)}{\partial \omega_1}, \dots,  \frac{ \partial \vec{v}(\theta_K,\omega_K)}{\partial \omega_K}  
    \end{bmatrix}. \label{eq:dAf}
\end{align}
\subsubsection{Unknown noise variance} When the noise power $\sigma^2$ is unknown, the FIM for estimating $\bm{\zeta}$ is given by: 
\begin{align}
    \vec{F}_{0}(\bm{\zeta})=\left[\begin{matrix}
       \vec{F}_{0}(\bm{\varphi}) & \vec{0} \nonumber \\
       \vec{0}   &  \frac{4NM_r}{\sigma^2}
    \end{matrix}\right],
\end{align}
which is a block diagonal matrix. Hence,  the  CRB matrix for the high precision data of both the known and unknown noise power cases is given by: 
\begin{equation}
    \vec{CRB}_{0}(\bm{\varphi}) = \vec{F}_{0}^{-1}(\bm{\varphi}). 
\end{equation}
\subsection{Cram{\'e}r-Rao Bound for One-Bit Data}
\subsubsection{Known noise variance} Making use of the property of Hadamard product: $(\vec{A}\vec{C})\odot (\vec{B}\vec{D})=(\vec{A}^T\oplus \vec{B}^T)^T(\vec{C}\oplus \vec{D})$, the FIM $\vec{F}_{0}(\bm{\varphi})$  for the high-precision data in \eqref{eq:F0} can be rewritten as:
\begin{equation}
    \vec{F}_{0}(\bm{\varphi})=\frac{2}{\sigma^2} \Re \big \{
    \vec{U}\vec{U}^H 
\big \}, \label{eq:29}
\end{equation}
where 
\begin{footnotesize}
\begin{align}
    \vec{U}=\begin{bmatrix} \dot{\vec{A}}_{\rm r}\oplus (\vec{V}\vec{B})+ \vec{A}_{\rm r} \oplus  (\dot{\vec{V}}_{\vec{\theta}}\vec{B}), \
     \vec{A}_{\rm r}\oplus (\dot{\vec{V}}_{\bm{\omega}}\vec{B}), \ \vec{A}_{\rm r}\oplus\vec{V}, \ j\vec{A}_{\rm r}\oplus \vec{V} \end{bmatrix}^H  \nonumber \\
     \in \mathbb{C}^{4K\times NM_r}. \label{eq:U}
\end{align}
\end{footnotesize}
The results presented in \cite{li2018bayesian,stoica2021cramer} have indicated the connection between the FIM  for one-bit quantizer $\vec{F}_{1}(\bm{\varphi})$ and that of the high-precision quantizer $\vec{F}_{0}(\bm{\varphi})$, from which  and \eqref{eq:29} we obtain the following expression for $\vec{F}_{1}(\bm{\varphi})$:
\begin{equation}
    \vec{F}_{1}(\bm{\varphi})=\frac{1}{\pi\sigma^2} (\vec{U}_{\rm R}\vec{\Lambda}_{\rm R} \vec{U}_{\rm R}^T + \vec{U}_{\rm I}\vec{\Lambda}_{\rm I}\vec{U}_{\rm I}^T),
\end{equation}
where $\vec{\Lambda}={\rm diag}(\begin{bmatrix} \lambda_1, \dots, \lambda_{NM_r}  \end{bmatrix})$. Let $\vec{h}={\rm vec}(\vec{H})$ and $\bm{\mu}=\frac{\sum_{k=1}^K b_k \vec{v}(\theta_k,\omega_k) \otimes \vec{a}_{\rm r}(\theta_k) -\vec{h}}{\sigma/\sqrt{2}}\in \mathbb{C}^{NM_r\times 1}$. Then, $\lambda_k$ is given by: 
\begin{equation}
    \lambda_k=G\left(\mu_{{\rm R},k}\right) + j G\left(\mu_{{\rm I},k}\right), 
\end{equation}
where the function $G(x)$ is defined as:
\begin{equation}
    G(x)=\left[\frac{1}{\Phi(x)} +\frac{1}{\Phi(-x)} \right]e^{-x^2}, \label{eq:Gx}
\end{equation}
with $\Phi(x)=\int_{-\infty}^{x} \frac{1}{\sqrt{2\pi}} e^{-\frac{t^2}{2}}dt$ being the cumulative distribution function (cdf) of the normal standard distribution.
The corresponding CRB for the one-bit measurements for the known $\sigma^2$ case is given by:
\begin{align}
 {\vec{CRB}_1}(\bm{\varphi})=\vec{F}_1^{-1}(\bm{\varphi}) .
\end{align}
We have proved in \cite{stoica2021cramer} that $0 < G(x)\leq 4$ and therefore $\vec{F}_1(\bm{\varphi}) \preceq \frac{2}{\pi}\vec{F}_0(\bm{\varphi})$. Moreover, combining the upper bounds for one-bit FIM in \cite{stoica2021cramer}, we have 
\begin{equation}
    \frac{2}{\pi}\vec{F}_{1,{\rm l}}(\bm{\varphi}) \preceq \vec{F}_1(\bm{\varphi}) \preceq \frac{2}{\pi}\vec{F}_0(\bm{\varphi}),  \label{eq:up_and_low}
\end{equation}
where $\vec{F}_{1,{\rm l}}(\bm{\varphi})=\frac{2}{\sigma^2}(\vec{U}_{\rm R} \bar{\vec{\Lambda}}_{\rm R}\vec{U}_{\rm R}^T + \vec{U}_{\rm I}\bar{\vec{\Lambda}}_{\rm I}\vec{U}_{\rm I}^T)$ with $\bar{\vec{\Lambda}}$ being a diagonal matrix with  its $k$-th diagonal element $\bar{\lambda}_k$ being:
\begin{equation}
    \bar{\lambda}_k=e^{-\mu_{{\rm R},k}^2} + je^{-\mu_{{\rm I},k}^2}. \label{eq:lambda}
\end{equation} Thus the one-bit CRB has the following lower and upper bounds:
\begin{equation}
    \frac{\pi}{2}\vec{F}_{1,{\rm l}}^{-1}(\bm{\varphi}) \succeq \vec{CRB}_1(\bm{\varphi}) \succeq \frac{\pi}{2}\vec{CRB}_0(\bm{\varphi}). 
\end{equation}
\subsubsection{Unknown noise variance} When the noise power $\sigma^2$ is unknown, the matrix $\vec{U}$ can be expanded as $\bar{\vec{U}}$ with the following expression:
\begin{align}
    \bar{\vec{U}}=\begin{bmatrix}\vec{U}^T \ -(\vec{x}-\vec{h})/ \sigma \end{bmatrix}^T\in \mathbb{C}^{(4K+1)\times NM_r}. \label{eq:Ubar}
\end{align}
Then the FIM for estimating $\bm{\zeta}$ is:
\begin{align}
    \vec{F}_1(\bm{\zeta})&=\left[ \begin{matrix} 
    \vec{F}_1(\bm{\varphi}) & \vec{F}_1({\bm{\varphi},\sigma}) \nonumber \\
    \vec{F}^{T}_1(\bm{\varphi},\sigma) & F_1(\sigma,\sigma)
    \end{matrix} \right] \nonumber \\
    &=\frac{1}{\pi\sigma^2} ( \bar{\vec{U}}_{\rm R}\vec{\Lambda}_{\rm R} \bar{\vec{U}}_{\rm R}^T + \bar{\vec{U}}_{\rm I}\vec{\Lambda}_{\rm I}\bar{\vec{U}}_{\rm I}^T), 
\end{align}
and the $\vec{CRB}_1(\bm{\varphi})$ for the unknown $\sigma^2$ case is given by:
\begin{equation}
    \vec{CRB}_1(\bm{\varphi})=\left[\vec{F}_{1}(\bm{\varphi}) - \vec{F}_1(\bm{\varphi},\sigma)F_1^{-1}(\sigma,\sigma)\vec{F}_1^T(\bm{\varphi},\sigma)\right]^{-1}.
\end{equation}
It can be easily verified that the inequality   $\vec{CRB}_1(\bm{\varphi}) \succeq \frac{\pi}{2}\vec{CRB}_0(\bm{\varphi})$  still holds for the unknown noise variance case. 
\subsection{Cram{\'e}r-Rao Bound for Mixed-ADC based Output} \label{sec:mixed-ADC}
\subsubsection{Known noise variance} 
We can observe from \eqref{eq:U} that the $k$th column of $\vec{U}$ represents the contribution of the $k$th measurement $x_k$ on the FIM of the high-precision outputs. Using the following definitions:
\begin{align}
    \vec{U}_0 &=\vec{U}{\rm diag}(\bm{\delta}\otimes \vec{1}_N), \nonumber \\
    \vec{U}_1 &=\vec{U}{\rm diag}(\bm{\bar{\delta}}\otimes \vec{1}_N),
\end{align}
we obtain the following expression for the mixed-ADC based FIM:
\begin{align}
    \vec{F}_{\rm m}(\bm{\varphi})=&\frac{2}{\sigma^2} \Re \left\{ 
    \vec{U}_0\vec{U}_0^H \right\} \nonumber \\
    &+ \frac{1}{\pi\sigma^2}( \vec{U}_{1,{\rm R}}\vec{\Lambda}_{\rm R}\vec{U}_{1,{\rm R}}^T + \vec{U}_{1,{\rm I}}\vec{\Lambda}_{\rm I}\vec{U}_{1,{\rm I}}^T), \label{eq:fimmixed}
\end{align}
where the first and  second terms of $\vec{F}_{\rm m}$, respectively, represent the contributions of the high-precision and one-bit measurements to the mixed-ADC based FIM. 
The  difference between the FIM for the mixed-ADC based outputs (i.e., $\vec{F}_{\rm m}(\bm{\varphi})$)  and that of the high precision outputs  (i.e., $\vec{F}_0(\bm{\varphi})$) can be expressed as:
\begin{align}
    (\vec{F}_0 -\vec{F}_{\rm m})^{-1}=\vec{F}_{0}^{-1} -\vec{F}_0^{-1}\left( \vec{F}_0^{-1}-\vec{F}_{\rm m}^{-1} \right)^{-1}\vec{F}_0^{-1}, \label{eq:dF}
\end{align}
where the dependence of $\vec{F}_0$ and $\vec{F}_m$ on $\bm{\varphi}$ is omitted to
simplify the notation; this dependence on $\bm{\varphi}$  will be reinstated when it becomes important. From \eqref{eq:dF}, we have that 
\begin{align}
    \vec{F}_0^{-1}-\vec{F}_{\rm m}^{-1}&=\left[ \vec{F}_0 -\vec{F}_0\left(\vec{F}_0 -\vec{F}_{\rm m} \right)^{-1}\vec{F}_0 \right]^{-1}. \label{eq:dF1}
\end{align}
Note that 
\begin{small}
\begin{align}
    \vec{F}_0-\vec{F}_{\rm m} &= \frac{2}{\sigma^2} \Re \left\{ 
    \vec{U}_1\vec{U}_1^H \right\}  -  \frac{1}{\pi\sigma^2}( \vec{U}_{1,{\rm R}}\vec{\Lambda}_{\rm R}\vec{U}_{1,{\rm R}}^T + \vec{U}_{1,{\rm I}}\vec{\Lambda}_{\rm I}\vec{U}_{1,{\rm I}}^T), \nonumber \\
    &  \succeq \left( 1- \frac{2}{\pi}\right)\bar{\vec{F}}_0(\vec{\bm{\varphi}}), \label{eq:Fml} 
\end{align}
\end{small} 
where $\bar{\vec{F}}_0(\vec{\bm{\varphi}}) \triangleq \frac{2}{\sigma^2}\Re \left\{\vec{U}_1\vec{U}_1^H \right\}$. Inserting \eqref{eq:Fml} into  \eqref{eq:dF1}, it is straightforward to obtain that
\begin{align}
    \vec{F}_{\rm m}^{-1} \succeq \vec{F}_0^{-1} +  \left[  \frac{\pi}{\pi-2}\vec{F}_0\bar{\vec{F}}_0^{-1} -\vec{I} \right]^{-1} \vec{F}^{-1}_0,
\end{align}
or, equivalently,
\begin{align}
    \vec{CRB}_{\rm m}(\bm{\varphi}) \succeq (\vec{I} + \bm{\Gamma}_{\rm l}) \vec{CRB}_0(\bm{\varphi}), \label{eq:lowerm}
\end{align} where $\bm{\Gamma}_{\rm l}\triangleq  \left[  \frac{\pi}{\pi-2}\vec{F}_0\bar{\vec{F}}_0^{-1} -\vec{I} \right]^{-1}$.   Combining the fact that $\vec{F}_0 \succ \bar{\vec{F}}_0$, $\bm{\Gamma}_{\rm l}$ is a positive-definite matrix corresponding to the loss paid for using one-bit ADC's. Moreover, $\bar{\vec{F}}_0^{-1}$ decreases as the number of one-bit ADC pairs (i.e. $M_{\rm r}-L$) increase; hence  $\vec{CRB}_{\rm m}(\bm{\varphi})$ increases, indicating inferior parameter estimation performance,  as the number of  one-bit ADC pairs increases.  Similarly, making use of the lower bound of the one-bit FIM in \eqref{eq:up_and_low}, we can get the following upper bound on $\vec{CRB}_{\rm m}(\bm{\varphi})$:
\begin{align}
    \vec{CRB}_{\rm m}(\bm{\varphi}) \preceq  (\vec{I} + \bm{\Gamma}_{\rm u})\vec{CRB}_0(\bm{\varphi}), \label{eq:uprm}
\end{align}
where  $\bm{\Gamma}_{\rm u}\triangleq  \left[  \vec{F}_0\left( \vec{F}_0 - \bar{\vec{F}}_{1,{\rm l}} \right)^{-1}-\vec{I} \right]^{-1}$ with 
\begin{align}
    \bar{\vec{F}}_{1,{\rm l}} = \vec{U}_{1,{\rm R}}\bar{\Lambda}_{\rm R}\vec{U}^T_{1,{\rm R}} +  \vec{U}_{1,{\rm I}}\bar{\Lambda}_{\rm I}\vec{U}^T_{1,{\rm I}},
\end{align}
and where $\bar{\bm{\Lambda}}$ is given in \eqref{eq:lambda}. 

\subsubsection{Unknown noise variance}
When $\sigma^2$ is unknown, the FIM for estimating $\bm{\zeta}$  is: 
\begin{align}
    \vec{F}_{\rm m}(\bm{\zeta})&=\left[ \begin{matrix} 
    \vec{F}_{\rm m}(\bm{\varphi}) & \vec{F}_{\rm m}({\bm{\varphi},\sigma}) \nonumber \\
    \vec{F}^{T}_{\rm m}(\bm{\varphi},\sigma) & F_{\rm m}(\sigma,\sigma)
    \end{matrix} \right] \nonumber \\
    &=\frac{2}{\sigma^2} \Re \left\{ \left[ \begin{matrix} 
    \vec{U}_0\vec{U}_0^H & \vec{0} \\
    \vec{0} & 2NL
    \end{matrix}
    \right]\right\} 
     \nonumber \\
    & + \frac{1}{\pi\sigma^2}( \bar{\vec{U}}_{1,{\rm R}}\vec{\Lambda}_{\rm R}\bar{\vec{U}}_{1,{\rm R}}^T + \bar{\vec{U}}_{1,{\rm I}}\vec{\Lambda}_{\rm I}\bar{\vec{U}}_{1,{\rm I}}^T), \label{eq:fimmixed_unknown}
\end{align}
where $\bar{\vec{U}}_1=\bar{\vec{U}}{\rm diag}(\bar{\bm{\delta}}\otimes \vec{1}_N)$. Similar to the one-bit case, the CRB matrix for estimating $\bm{\varphi}$ using  the mixed-ADC based measurements is given by: 
\begin{equation}
    \vec{CRB}_{\rm m}(\bm{\varphi})=\left[\vec{F}_{\rm m}(\bm{\varphi}) - \vec{F}_{\rm m}(\bm{\varphi},\sigma)F_{\rm m}^{-1}(\sigma,\sigma)\vec{F}_{\rm m}^T(\bm{\varphi},\sigma)\right]^{-1}.
\end{equation}
Also, it can be readily checked that the lower bound shown in \eqref{eq:lowerm} still holds for this unknown noise variance case. 
\section{Mixed-ADC based parameter estimation }
Consider first  the maximum likelihood (ML) estimator, which is theoretically appealing due to its desirable properties including consistency and asymptotic efficiency.  We assume that both $\vec{e}_0\in \mathbb{C}^{NL\times 1}$ and  $\vec{e}_1\in \mathbb{C}^{N(M_{\rm r}-L)\times 1}$ are  circularly symmetric complex-valued white Gaussian noise with i.i.d. $\mathcal{CN}(0,\sigma^2)$ entries. The negative log-likelihood function for the model described in \eqref{eq:vecmodelnew} is given by: 
\begin{footnotesize}
\begin{align}
    &-\ln{L(\bm{\zeta})}=-\ln{L_0(\bm{\zeta})} - \ln{L_1(\bm{\zeta})} \nonumber \\
    &=  -\sum_{n=1}^{N(M_r-L)}\ln\left(\Phi\left(y_{\rm 1,R}(n)\frac{\Re\lbrace \sum_{k=1}^K b_k a_{1,n}(\theta_k,\omega_k) \rbrace  -{h}_{\rm 1,R}(n) }{\sigma / \sqrt{2}} \right) \right) \nonumber \\
     &\quad -\sum_{n=1}^{N(M_r-L)}\ln\left(\Phi\left(y_{\rm 1,I}(n)\frac{\Im\lbrace \sum_{k=1}^K b_k{a}_{1,n}(\theta_k,\omega_k) \rbrace  -{h}_{\rm 1,I}(n) }{\sigma / \sqrt{2}} \right) \right) \nonumber \\
    & \quad + \frac{\Vert \vec{y}_0 -\sum_{k=1}^Kb_k\vec{a}_0(\theta_k,\omega_k) \Vert^2}{\sigma^2} + NL\ln{\sigma^2} + NL\ln{\pi}, \label{eq:likelihood}
\end{align}
\end{footnotesize}where $a_{1,n}(\theta_k,\omega_k)$ represents the $n$-th element of the vector $\vec{a}_1(\theta_k,\omega_k)$. Let $\eta=1/\sigma$ and $\bm{\beta}=\eta \vec{b}$. Then the negative log-likelihood function can be rewritten as:
\begin{footnotesize}
\begin{align}
    &-\ln{L(\bm{\beta},\bm{\omega},\bm{\theta},\eta)} \nonumber \\ &=-\sum_{n=1}^{N(M_r-L)}\ln\left(\Phi\left(y_{\rm 1,R}(n){\sqrt{2}}  \left[\Re\lbrace \sum_{k=1}^K \beta_k {a}_{1,n}(\theta_k,\omega_k) \rbrace  -\eta {h}_{\rm 1,R}(n) \right]  \right) \right) \nonumber \\
     &\quad -\sum_{n=1}^{N(M_r-L)}\ln\left(\Phi\left(y_{\rm 1,I}(n){\sqrt{2}} \left[\Im\lbrace \sum_{k=1}^K \beta_k {a}_{1,n}(\theta_k,\omega_k) \rbrace  -\eta{h}_{\rm 1,I}(n) \right] \right) \right) \nonumber \\
    &\quad  + \Vert \eta \vec{y}_0 -\sum_{k=1}^K \beta_k\vec{a}_0(\theta_k,\omega_k) \Vert^2 - NL\ln{\eta^2} + NL\ln{\pi}. \label{eq:mle}
\end{align}
\end{footnotesize}
The negative log-likelihood function has a complicated form and is highly nonlinear and nonconvex with regard to $\bm{\theta}$ and $\bm{\omega}$. Therefore, the global minimization of $-\ln{L(\bm{\beta},\bm{\omega},\bm{\theta},\eta)}$ is a challenging problem. For given $\bm{\theta}$ and $\bm{\omega}$, however, the above optimization problem is convex with regard to $\bm{\beta}$ and $\eta$ and hence  can be globally and efficiently solved using, e.g., the Newton method.  Consequently, the ML estimator can be directly implemented as follows:  
\begin{enumerate}
\item[A.] Perform a $2K$-dimensional exhaustive coarse search on the angular and Doppler  spaces to find coarse estimates of both $\bm{\theta}$ and $\bm{\omega}$.
\item[B.] Determine the corresponding optimal estimates of $\bm{\beta}$ and $\eta$ for given estimates of    $\bm{\theta}$ and $\bm{\omega}$.
\item[C.]  Refine the results by using an efficient fine search method over  narrow intervals near the coarse angle and Doppler estimates. 
\end{enumerate}
Regarding Step A, we consider a direct grid-based implementation. Assume that the angular and Doppler domains are, respectively, grided into $K_{\theta}$ and $K_{\omega}$ points. Then the optimization problem of estimating $\bm{\beta}$ and $\eta$ for given estimates of  $\bm{\theta}$ and $\bm{\omega}$ needs to  be solved  $(K_{\theta}K_{\omega})^K$ times and  the parameters with the minimum negative log-likelihood value are chosen as  their estimates.  As the number of targets increases, the  complexity required by step A becomes computationally prohibitive. We  present next an efficient  strategy to realize the ML estimation.  
\subsection{mLIKES for Sparse Parameter Estimation}
We uniformly discretize the continuous angular and Doppler space into $K_{\theta}$ and $K_{\omega}$ grid points, respectively. {To attain high resolution, the typical choices of $K_{\theta}$ and $K_{\omega}$ are $4M_{\rm r}\leq K_{\theta} \leq 10M_{\rm r}$ and $4N\leq K_{\omega} \leq 10N$, respectively. The number of targets, i.e., $K$, is usually much smaller than $K_{\theta}K_{\omega}$. Therefore we can formulate the data model in \eqref{eq:vecmodel},  at least approximately, as the following sparse linear model:} 
\begin{equation}
    \vec{x}=\vec{A}\bm{\alpha} + \vec{e} \in \mathbb{C}^{NM_r}, \ \vec{A}\in \mathbb{C}^{NM_r \times K_{\theta}K_{\omega}},  \label{eq:approx1}
\end{equation}
where $\vec{A}=\begin{bmatrix}\vec{a}(\tilde{\theta}_1,\tilde{\omega}_1),\dots, \vec{a}(\tilde{\theta}_{k_{\theta}},\tilde{\omega}_{k_{\omega}}),  \dots, \vec{a}(\tilde{\theta}_{K_{\theta}},\tilde{\omega}_{K_{\omega}}) \end{bmatrix}$,  with    $\vec{a}(\tilde{\theta}_{k_{\theta}},\tilde{\omega}_{k_{\omega}})$ defined as:
\begin{footnotesize}
\begin{equation}
    \vec{a}(\tilde{\theta}_{k_{\theta}},\tilde{\omega}_{k_{\omega}})= \left[ \left(\vec{C}^T \otimes \vec{I}_{M_{\rm r}}\right) \left( \vec{a}_{\rm t}(\tilde{\theta}_{k_{\theta}}) \otimes \vec{a}_{\rm r}(\tilde{\theta}_{k_{\theta}}) \right) \right] \odot \left[\vec{d}(\tilde{\omega}_{k_\omega}) \otimes \vec{1}_{M_r} \right], 
\end{equation} 
\end{footnotesize}and $\bm{\alpha} \in \mathbb{C}^{K_{\theta}K_{\omega}}$ is an  unknown sparse vector with many zero elements.  The non-zero elements of $\bm{\alpha}$ correspond to the target parameters.  It follows from \eqref{eq:vecmodel} and \eqref{eq:approx1} that \eqref{eq:vecmodelnew} can also be rewritten as:
\begin{align}
    \vec{y}_0&=\vec{A}_{0}\bm{\alpha} + \vec{e}_0 \in \mathbb{C}^{NL\times 1}, \nonumber \\
    \vec{y}_1&={\rm signc}(\vec{A}_1\bm{\alpha}-\vec{h}_1 +\vec{e}_1) \in \mathbb{C}^{N(M_r-L)\times 1}, \label{eq:sparsemodel}
\end{align}
where $\vec{A}_0 \triangleq \vec{A}\left[ \bm{\delta}\otimes \vec{1}_N ,:\right]\in \mathbb{C}^{NL\times K_\theta K_\omega}$ and $\vec{A}_1=\vec{A}\left[ \bar{\bm{\delta}}\otimes \vec{1}_N ,:\right] \in \mathbb{C}^{N(M_{\rm r}-L)\times K_\theta K_\omega}$, respectively,  correspond to the dictionary matrices of the high-precision measurements and the one-bit measurements.  To enhance the robustness of  mLIKES, we allow the noise of the high-precision outputs and that of the one-bit outputs to have different powers (namely $\vec{e}_0 \sim \mathcal{CN}(\vec{0},\sigma_0^2\vec{I}_{NL})$ and $\vec{e}_1\sim \mathcal{CN}\left(\vec{0},\sigma_1^2\vec{I}_{N(M_r-L)}\right)$). Let $\eta_1=1/ \sigma_1$. From  \eqref{eq:sparsemodel}, the negative log-likelihood  function of the one-bit measurements is:
\begin{footnotesize}
\begin{align}
    -\ln L_1(\bm{\alpha},\eta_1)= & - \sum_{n=1}^{N(M_r-L)} \ln \Phi  \left( y_{1,{\rm R}}(n) {\sqrt{2}}{\eta_1} \Re \left[ \vec{a}_{1,n}^T\bm{\alpha}-h_{1}(n)\right] \right)  \nonumber \\
    &- \sum_{n=1}^{N(M_r-L)} \ln \Phi  \left( y_{1,{\rm I}}(n) {\sqrt{2}} {\eta_1} \Im \left[ {\vec{a}}_{1,n}^T\bm{\alpha}-h_{1}(n)\right] \right), 
\end{align}
\end{footnotesize}
where $\vec{a}_{1,n}$ is the $n$-th column of $\vec{A}_{1}^T$.

Then, taking into account the sparse property of the signal model, we  minimize the following objective function for sparse parameter estimation:
\begin{align}
    \Psi (\bm{\alpha},\eta_1,\sigma_0,\vec{p})= -\ln L_{1}(\bm{\alpha},\eta_1) &+ \frac{1}{\sigma_0^2} \Vert \vec{A}_{0}\bm{\alpha}- \vec{y}_0 \Vert^2 \nonumber \\  
    &+ \sum_{k=1}^{K_{\theta}K_\omega} \frac{|\alpha_k|^2}{p_k} + \ln |\vec{R}|,\label{eq:objective}
\end{align}
where  
\begin{align}
    \vec{R}&=\left[ \begin{matrix}
       \vec{A}_0  \\
       \vec{A}_1
    \end{matrix}\right]\vec{P} \left[ 
    \begin{matrix}
       \vec{A}_0^H & \vec{A}_1^H
    \end{matrix}
    \right] + \left[ \begin{matrix} 
     \sigma_0^2 \vec{I}_{NL} & \vec{0}  \\
    \vec{0} & \frac{1}{\eta_1^2}\vec{I}_{N(M_r-L)}
    \end{matrix}\right] \nonumber \\ 
    &= \tilde{\vec{A}}\vec{P}\tilde{\vec{A}}^H + \bm{\Sigma} \label{eq:conv}
\end{align}
and
\begin{align}
    \tilde{\vec{A}} & \triangleq \begin{bmatrix} \vec{A}_0^T & \vec{A}_1^T \end{bmatrix}^T  \nonumber \\
    \bm{\Sigma}&={\rm diag}\left( \begin{bmatrix} \sigma_0^2\vec{1}_{NL}^T, & 1/ \eta_1^2 \vec{1}_{N(M_r-L)}^T \end{bmatrix}^T \right) \nonumber \\
    \vec{P}&={\rm diag}(\vec{p}), \ \vec{p}=\begin{bmatrix}p_1,\dots, p_{K_\theta K_{\omega}} \end{bmatrix}^T.
\end{align}
 
The first term in \eqref{eq:objective}, i.e., $-\ln L_1(\bm{{\alpha}},\eta_1)$,  is the fitting term of the one-bit measurements; the second term $\frac{1}{\sigma_0^2}\Vert{\vec{A}}_{0}{\bm{\alpha}}-\vec{y}_{0} \Vert^2$ represents the least-squares fitting term of the high-precision measurements; and $\ln |\vec{R}|$ serves as the sparsity enforcing term. Note that the summation of the last three terms in \eqref{eq:objective} is an augmented form of the cost function of the conventional LIKES algorithm (see \cite{stoica2014weighted} for more details), and hence mLIKES reduces to LIKES when there are no one-bit measurements.  Note that $\ln|\vec{R}|$ is a concave function of $\vec{p}$, $\sigma_0^2$ and $\frac{1}{\eta_1^2}$, and we use the majorization-minimization (MM)  technique \cite{stoica2004cyclic} to simply it to attain enhanced computational efficiency.   
\par Given the estimates of $\vec{p},\sigma_0$, $\eta_1$ at the $m$-th MM iteration and $\hat{\vec{R}}^m$ constructed by \eqref{eq:conv},  a majorizing function for $\ln{| \vec{R} |}$ can be easily constructed by its first-order Taylor expansion:
\begin{align}
    \ln{| \vec{R} |} \leq \sum_{k=1}^{K_\theta K_\omega} w_k p_k + \bar{w}_0 \sigma_0^2 +  \frac{\bar{w}_1}{\eta_1^2}, \label{eq:majorlnR}
\end{align}
where 
\begin{align}
    w_k&=\tilde{\vec{a}}_{k}^H (\hat{\vec{R}}^m)^{-1} \tilde{\vec{a}}_k, \ k=1,\dots, K_\theta K_\omega, \nonumber \\
   \bar{w}_0 &= \vec{1}^T_{NL}\tilde{\vec{p}} [\bm{\delta}\otimes \vec{1}_N], \nonumber \\  \bar{w}_1 &=\vec{1}^T_{N(M_r-L)} \tilde{\vec{p}}[\bar{\bm{\delta}}\otimes \vec{1}_N]  \ {\rm with} \ \tilde{\vec{p}}={\rm diag}((\hat{\vec{R}}^m)^{-1}),  \label{eq:computew}
\end{align}
with $\tilde{\vec{a}}_k$ being the $k$-th column of $\tilde{\vec{A}}$.  Making use of \eqref{eq:majorlnR},  we can obtain a majorizing function for \eqref{eq:objective} at the $(m+1)$-th MM iteration as follows: 
\begin{equation}
    Q^{m+1}(\bm{\alpha},\vec{p},\sigma_0,\eta_1)= -\ln{L_1(\bm{\alpha},\eta_1)}  + \psi^{m+1}(\bm{\alpha},\vec{p},\sigma_0,\eta_1), \label{eq:Gfista} 
\end{equation}
where
\begin{align}
    \psi^{m+1}(\bm{\alpha},\vec{p},\sigma_0,\eta_1) \triangleq &  
    \frac{1}{\sigma_0^2} \Vert \vec{A}_0 \vec{b} -\vec{y}_0 \Vert^2 + \sum_{k=1}^{K_{\theta}K_\omega}  \frac{|\alpha_k|^2}{p_k} \nonumber \\
    & +\sum_{k=1}^{K_\theta K_\omega} w_kp_k +  \bar{w}_0 \sigma_0^2 +  \frac{\bar{w}_1}{\eta_1^2}.
\end{align}
The minimization of $Q^{m+1}(\bm{\alpha},\vec{p},\sigma_0,\eta_1)$ is still a highly nonlinear and non-convex optimization problem and  hard to solve. To mitigate the difficulty,  we again adopt the MM technique, referred to as the inner MM iteration, to iteratively minimize \eqref{eq:Gfista}.  Let $f(x)=-\ln{\Phi(x)}$. The analysis of $f(x)$ in \cite{ren2019sinusoidal} demonstrates that for all ${x}\in \mathbb{R}$, $f(x)$ has a bounded second-order derivative: 
\begin{equation}
   0<     f^{''}(x) < 1. \label{eq:secd}
\end{equation}
The following inequality follows from the Lagrange’s mean value theorem and the inequality in \eqref{eq:secd}:
\begin{equation}
    \frac{|f^{'}(x)-f^{'}(y)|}{|x-y|} = |f^{''}(\xi)| < 1 \ {\rm for \ every} \ x,y\in\mathbb{R}, \label{eq:Lipschitz}
\end{equation}
where $x\leq \xi \leq y$. The inequality in \eqref{eq:Lipschitz} implies that $f(x)$ is a continuously differentiable function with Lipschitz constant equal to 1.  Then, the following inequality holds  for any $x,y\in \mathbb{R}$
\begin{align}
    f(x) &\leq f(y) + f^{'}(y)(x-y) + \frac{1}{2}(x-y)^2 \nonumber \\
    &= \frac{1}{2}\left[x-\left(y - \nabla f(y)  \right) \right]^2 +{\rm const}. \label{ineq:Laptiz}
\end{align} Define $\bm{\gamma}=\begin{bmatrix}\gamma (1),\dots, \gamma(N(M_r-L)) \end{bmatrix}^T$ with 
\begin{align}
    \gamma_{\rm R}(n)&= y_{\rm 1,R}{\sqrt{2}}{\eta_1}  \Re\left[\vec{a}_{1,n}\bm{\alpha}-h_{1}(n)\right] \nonumber \\
    \gamma_{\rm I}(n)&= y_{\rm 1,I}{\sqrt{2}}{\eta_1}\Im\left[\vec{a}_{1,n}\bm{\alpha}-h_{1}(n)\right]. \label{eq:def1}
\end{align}
Using this notation,  $-\ln{L_1}(\bm{\alpha},\eta_1)$ can be rewritten as:
\begin{equation}
    -\ln{L_1(\bm{\alpha},\eta_1)}= \sum_{n=1}^{N(M_r-L)} f\left(\gamma_{\rm R}(n)\right) + \sum_{n=1}^{N(M_r-L)} f\left(\gamma_{\rm I}(n)\right). 
\end{equation}
Making use of \eqref{ineq:Laptiz}, a majorizing function for $-\ln{L_1}(\bm{\alpha},\eta_1)$ can be constructed as follows:
\begin{align}
    -\ln{L_1}(\bm{\alpha},\eta_1) & \leq \frac{1}{2} \Vert \bm{\gamma} -  \left( \hat{\bm{\gamma}}^{m+1}_i - \nabla f(\hat{\bm{\gamma}}^{m+1}_i) \right) \Vert^2 + 
     {\rm const},
\end{align}
where $\hat{\gamma}_{{\rm R/I},i}^{m+1}(n)$  is the estimate of $\gamma_{\rm R/I}(n)$ at the $i$-th inner MM iteration performed at the $(m+1)$-th outer MM iteration.  Combined with $\psi(\bm{\alpha},\vec{p},\sigma_0,\eta_1)$,  a majorizing function for $Q^{m+1}(\bm{\alpha},\bm{p},\sigma_0,\eta_1)$ at the $(i+1)$-th iteration is given as follows:
\begin{align}
    Q^{m+1}(\bm{\alpha},\bm{p},\sigma_0,\eta_1) \leq & \frac{1}{2} \Vert \bm{\gamma} -  \left( \hat{\bm{\gamma}}^{m+1}_i - \nabla f(\hat{\bm{\gamma}}^{m+1}_i) \right) \Vert^2  \nonumber \\
    & + \psi^{m+1}(\bm{\alpha},\vec{p},\sigma_0,\eta_1) + {\rm const}. \label{eq:mmjorization}
\end{align}
Inserting \eqref{eq:def1} into \eqref{eq:mmjorization} leads to the minimization of the following optimization criterion: 
\begin{align}
     \bar{\Psi}(\bm{\alpha},\vec{p},\sigma_0,\eta_1)=& \eta_1^2 \Vert\vec{A}_{1}\bm{\alpha}-\vec{h}_1-\frac{1}{\sqrt{2}\eta_1} \vec{g}^{m+1}_i \Vert^2 \nonumber \\
     & + \psi^{m+1}(\bm{\alpha},\vec{p},\sigma_0,\eta_1). \label{eq:objfista}
\end{align}
where $\vec{g}^{m+1}_i=\begin{bmatrix}g^{m+1}_i(1),\dots,g^{m+1}_i(N(M_{\rm r}-L))\end{bmatrix}^T$,
\begin{align}
    g^{m+1}_i(n)&=y_{1,{ \rm R}}(n) u_{{\rm R},i}^{m+1}(n) + j y_{1,{\rm I}}(n) u_{{{\rm I}},i}^{m+1}(n), \nonumber \\ 
    u^{m+1}_{{\rm R/I},i}(n)&=\hat{\gamma}^{m+1}_{{\rm R/I},i}(n)-\nabla f(\hat{\gamma}^{m+1}_{{\rm R/I},i}(n)),   \nonumber \\
    \hat{\gamma}^{m+1}_{{\rm R},i}(n)&= y_{1,{\rm R}} \sqrt{2} \hat{\eta}^{m+1}_i \Re \left[{\vec{a}}_{1,n}^T \hat{\bm{\alpha}}^{m+1}_i -h_{1}(n) \right], \nonumber
\end{align}
and
\begin{equation}
      \hat{\gamma}^{m+1}_{{\rm I},i}(n)= y_{1,{\rm I}} \sqrt{2} \hat{\eta}^{m+1}_i \Im \left[ \vec{a}_{1,n}^T \hat{\bm{\alpha}}^{m+1}_i -h_{1}(n) \right]. \label{eq:g_fista}
\end{equation}
The minimization of \eqref{eq:objfista} can be achieved by using a blockwise cyclic algorithm, which alternatingly minimizes \eqref{eq:objfista} with respect to $\bm{\alpha},\vec{p},\sigma_0$ and $\eta_1$. 
\subsubsection{Updating  $\vec{\alpha}$} 
Let 
\begin{equation}
    \bar{\vec{y}}_{1,i}^{m+1}=\vec{h}_1 + \frac{1}{\sqrt{2}\hat{\eta}^{m+1}_{1,i}}\vec{g}^{m+1}_i.
\end{equation}
Ignoring the terms independent of $\bm{\alpha}$ yields the following iterative scheme: 
\begin{footnotesize}
\begin{align}
     &\min_{\bm{\alpha}} (\hat{\eta}_{1,i}^{m+1})^2 \Vert \vec{A}_1\bm{\alpha} -\bar{\vec{y}}_{1,i}^{m+1} \Vert^2 + \frac{\Vert \vec{A}_0\vec{b} -\vec{y}_0 \Vert^2}{(\hat{\sigma}_{0,i}^{m+1})^2}  + \sum_{k=1}^{K_\theta K_{\omega}} \frac{|\alpha_k|^2}{\hat{p}_{k,i}^{m+1}},
\end{align} 
\end{footnotesize}
or equivalently,
\begin{footnotesize}
\begin{align}
   \min_{\bm{\alpha}} (\tilde{\vec{y}}^{m+1}_{i} - \tilde{\vec{A}}\bm{\alpha})^H (\hat{\vec{\Sigma}}^{m+1}_i)^{-1} (\tilde{\vec{y}}^{m+1}_i- \tilde{\vec{A}}\bm{\alpha}) + \sum_{k=1}^{K_fK_{\theta}} \frac{|\alpha_k|^2}{\hat{p}_{k,i}^{m+1}}, \label{eq:minb}
\end{align} 
\end{footnotesize}
where 
\begin{align}
    \tilde{\vec{y}}^{m+1}_i &=\begin{bmatrix} \vec{y}_0^T & (\bar{\vec{y}}_{1,i}^{m+1})^T \end{bmatrix}^T \nonumber \\
    \hat{\vec{\Sigma}}_i^{m+1}&= \left[ \begin{matrix}
    (\hat{\sigma}_{0,i}^{m+1})^2 \vec{I}_{NL} & \vec{0} \\
    \vec{0}  &  1/(\hat{\eta}_{1,i}^{m+1})^2\vec{I}_{N(M_r-L)}  
    \end{matrix}
    \right].
\end{align}
Lemma 3 in \cite{stoica2014weighted} indicates that minimizing \eqref{eq:minb} yields:
\begin{equation}
\hat{\bm{\alpha}}^{m+1}_{i+1}=\hat{\vec{P}}^{m+1}_i\tilde{\vec{A}}^H\left(\hat{\vec{R}}^{m+1}_i\right)^{-1}\tilde{\vec{y}}^{m+1}_{i}, \label{eq:updat_alpha}
\end{equation}
and the minimum value is 
\begin{equation}
    (\tilde{\vec{y}}^{m+1}_i)^H ( \hat{\vec{R}}^{m+1}_i )^{-1} \tilde{\vec{y}}_i^{m+1}. 
\end{equation}
\subsubsection{Updating $\vec{p}$} 
To update $\vec{p}$ at the $(i+1)$-th MM iteration, we need to solve the following subproblem:
\begin{equation}
\hat{\vec{p}}_{i+1}^{m+1} =\arg \min_{\vec{p}}\sum_{k=1}^{K_{\theta}K_{\omega}} \frac{|\hat{\alpha}_{k,i+1}^{m+1}|^2}{p_k} + \sum_{k=1}^{K_\theta K_{\omega}} w_k p_k. 
\end{equation}
It is easy to check that the solution to the above optimization problem is:
\begin{equation}
    \hat{p}_{k,i+1}^{m+1} = |\hat{\alpha}_{k,i+1}^{m+1}|/ \sqrt{w_k}.   \label{eq:update_p}
\end{equation}
\subsubsection{Updating $\sigma_0$ and $\eta_1$} 
Ignoring the terms independent of $\sigma_0$ yields the following subproblem:
\begin{equation} 
 \hat{\sigma}_{0,i+1}^{m+1} = \arg \min_{\sigma_0}\frac{1}{\sigma_0^2} \Vert\vec{A}_0\hat{\bm{\alpha}}^{m+1}_{i+1} -\vec{y}_0\Vert^2 + \bar{w}_0 \sigma_0^2.
\end{equation}
The solution to the above optimization problem is:
\begin{equation}
    \hat{\sigma}_{0,i+1}^{m+1}= \frac{\sqrt{\Vert \vec{A}_0\hat{\bm{\alpha}}^{m+1}_{i+1} -\vec{y}_0 \Vert}}{\bar{w}_0^{1/4}}.  \label{eq:update_sigma}
\end{equation}
Similarly, we can update $\eta_1$ by solving the following subproblem:
\begin{align}
    \hat{\eta}_{1,i+1}^{m+1} & = \arg \min_{\eta_1} \eta_1^2  \Vert \vec{A}_1\hat{\bm{\alpha}}^{m+1}_{i+1} - \vec{h}_1 -\frac{1}{\sqrt{2}\eta_1}\vec{g}^{m+1}_i \Vert^2 +  \frac{\bar{w}_1}{\eta_1^2}. \label{eq:update_eta}
\end{align}
The optimization problem in \eqref{eq:update_eta} contains just one variable and is obviously convex with respective to $\eta_1$.  Therefore, the updating of $\eta_1$  can be efficiently implemented by minimizing the above objective function by using, for example, the MATLAB fminbnd function. 

{ Note that the monotonicity property of the mLIKES algorithm
is guaranteed since
\begin{align}
\Psi(\hat{\bm{\alpha}}^m,\hat{\eta}_1^m,\hat{\sigma}_0^m,\hat{\vec{p}}^m)&=Q^{m+1}(\hat{\bm{\alpha}}^m,\hat{\eta}_1^m,\hat{\sigma}_0^m,\hat{\vec{p}}^m) \label{eq:eq} \\
&\geq Q^{m+1}(\hat{\bm{\alpha}}^{m+1},\hat{\eta}_1^{m+1},\hat{\sigma}_0^{m+1},\hat{\vec{p}}^{m+1}) \label{ineq:a}\\
& \geq \Psi(\hat{\bm{\alpha}}^{m+1},\hat{\eta}_1^{m+1},\hat{\sigma}_0^{m+1},\hat{\vec{p}}^{m+1}) \label{eq:ineq}
\end{align}
Both the equality in \eqref{eq:eq} and the inequality in \eqref{eq:ineq} follow from the property of the majorizing function. The inequality in \eqref{ineq:a} comes from the minimization of $Q^{m+1}(\bm{\alpha},\eta_1,\sigma_0,\vec{p})$ by using the inner MM iterations. 
Therefore mLIKES is guaranteed a local convergence.}

\textit{Remark 1:} Observe from \eqref{eq:mmjorization} that at the $(i+1)$-th inner MM iteration, we approximate $Q^{m+1}(\bm{\alpha},\vec{p},\sigma_0,\eta_1)$  by replacing $-\ln{L_1(\bm{\alpha},\eta_1)}$ with its linear approximation at $\hat{\bm{\gamma}}^{m+1}_i$, which is also the basic idea of the proximal gradient  descent algorithm \cite{bregman1967relaxation,beck2009fast}. It was shown in \cite{nesterov1983method,beck2009fast} that the Nesterov acceleration technique can speed up the  proximal gradient descent to attain an optimal convergence rate. Herein we also adopt the idea of the Nesterov acceleration technique to accelerate the convergence rate of the inner MM iterations. Instead of  approximating $-\ln{L_1(\bm{\alpha},\eta_1)}$ at the point $\hat{\bm{\gamma}}^{m+1}_i$, we can make the approximation at an extrapolated point $\tilde{\hat{\bm{\gamma}}}^{m+1}_i$, which  linearly combines the two previous  points $\lbrace \hat{\bm{\gamma}}^{m+1}_i, \hat{\bm{\gamma}}^{m+1}_{i-1} \rbrace$:
\begin{align}
    \tilde{\hat{\bm{\gamma}}}^{m+1}_{i} = \hat{\bm{\gamma}}^{m+1}_{{\rm R/I},i} + \left( \frac{t_{i}-1}{t_{i+1}}\right)(\hat{\gamma}^{m+1}_{{\rm R/I},i} - \hat{\gamma}^{m+1}_{{\rm R/I},i-1}), \label{eq:76}
\end{align}
with $t_{i+1}=\frac{1+\sqrt{1+4t_{i}^2}}{2}$ and $t_0=1$. At the $(i+1)$-th iteration, the optimization problem described in \eqref{eq:mmjorization} can be replaced by the following counterpart:
\begin{align}
    \min_{\bm{\alpha},\vec{p},\sigma_0,\eta_1} \eta_1^2 \Vert \bm{\gamma} - \left( \tilde{\hat{\bm{\gamma}}}^{m+1}_i - \nabla f(\tilde{\hat{\bm{\gamma}}}^{m+1}_i) \right) \Vert^2 + \psi^{m+1}(\bm{\alpha},\vec{p},\sigma_0,\eta_1). \label{eq:ffs} 
\end{align}
It follows from \eqref{eq:mmjorization} and  \eqref{eq:objfista} that  \eqref{eq:ffs} can be reformulated as:
\begin{equation}
     \min_{\bm{\alpha},\vec{p},\sigma_0,\eta_1} \eta_1^2 \Vert\vec{A}_{1}\bm{\alpha}-\vec{h}_1-\frac{1}{\sqrt{2}\eta_1} \tilde{\vec{g}}^{m+1}_i \Vert^2 + \psi^{m+1}(\bm{\alpha},\vec{p},\sigma_0,\eta_1), \label{eq:objfista1} 
\end{equation}
where  
\begin{align}
    \tilde{\vec{g}}^{m+1}_i= \vec{g}^{m+1}_i +\frac{t_i-1}{t_{i+1}}\left(\vec{g}^{m+1}_i - \vec{g}^{m+1}_{i-1} \right). \label{eq:fistag}
\end{align}
Note that the optimization problem in \eqref{eq:objfista1} has the same form as the one in \eqref{eq:objfista} except that $\vec{g}^{m+1}_i$ in \eqref{eq:objfista} takes the form in \eqref{eq:fistag}. The updating formulas for $\lbrace \bm{\alpha},\vec{p},\sigma_0, \eta_1 \rbrace$ have similar forms as
those in \eqref{eq:updat_alpha}, \eqref{eq:update_p}, \eqref{eq:update_sigma} and \eqref{eq:update_eta}. 

We summarize the detailed steps of the proposed  mLIKES  in Algorithm 1, where Steps $8 - 13$ correspond to the inner MM iterations. Our extensive  numerical simulations show that the inner MM iterations can typically converge within a few (e.g., 3) iterations while the number of outer MM iterations exceeds 5. 

\textit{Remark 3:} Compared with the inner updating steps of the conventional LIKES for the high-precision data (see \cite{stoica2012spice,stoica2014weighted} for details),  we find that the term  $\bar{\vec{y}}^{m+1}_{1,i}$  can be interpreted as an estimate of the high-precision data  $\vec{y}_1$. In each inner MM iteration, mLIKES reconstructs an estimate (i.e., $\bar{\vec{y}}^{m+1}_{1,i}$) of the high-precision received signal (i.e., $\vec{y}_1)$ using the most recently updated estimates from the previous step. Together with using the high-precision measurements $\vec{y}_0$, mLIKES updates the sparse vector $\bm{\alpha}$  and $\vec{p}$  by using  the same formulas as the LIKES algorithm for the high-precision counterpart (see \cite{stoica2014weighted}).  In other words, by using the  MM technique, the sparse parameter estimation problem for the mixed-ADC based model in \eqref{eq:sparsemodel}  can be approximately and iteratively solved by making use of a high-precision model with the high-precision data $\tilde{\vec{y}}_{i}^{m+1}$ given by:
\begin{equation}
    \tilde{\vec{y}}_i^{m+1}=\tilde{\vec{A}}\bm{\alpha} + \tilde{\vec{e}},
\end{equation}
where the mean and covariance matrix of the noise vector $\bar{\vec{e}}$ are 0 and $\hat{\bm{\Sigma}}^{m+1}_i$,  respectively. 
\begin{small}
\begin{table}[htb]
\renewcommand\arraystretch{1.0}
\centering
\begin{tabular}{l}
\toprule
\textbf{Algorithm 1:} mLIKES \\
\midrule
\textbf{Input:} The mixed-ADC measurements: $\vec{Y}$ (or $\vec{y}_0$ and $\vec{y}_1$); \\
\quad \quad \quad  Thresholds adopted by one-bit ADC: $\vec{h}_1$. \\
\textbf{Procedure:}\\
\quad 1:  Initialize $\hat{\bm{\alpha}}_0$, $\hat{\vec{p}}_0$, $\hat{\sigma}_{0,0}$ and $\hat{\eta}_{1,0}$; $m=0$.\\
\quad 2: \textbf{repeat}\\
\quad 3: \quad Update $\hat{\vec{R}}^{m}$ with $\hat{\vec{p}}^{m}, \hat{\sigma}_0^{m}$ and $\hat{\eta}_1^{m}$ using \eqref{eq:conv}.\\
\quad 4: \quad Compute $\lbrace w_k \rbrace_{k=1}^{K_\theta K_\omega},\bar{w}_0$ and $\bar{w}_1$ using \eqref{eq:computew}. \\
\quad 5: \quad Compute ${\vec{g}}^{m+1}_0$  using \eqref{eq:g_fista}; $\tilde{\vec{g}}^{m+1}_0={\vec{g}}^{m+1}_0$;   \\
\quad 6: \quad $\hat{\vec{R}}^{m+1}_0=\hat{\vec{R}}^m$; $i=0$; $t_0=1$.  \\
\quad 7: \quad \textbf{repeat}\\
\quad 8: \quad \quad Update $\hat{\bm{\alpha}}^{m+1}_{i+1}$ and $\hat{\vec{p}}^{m+1}_{i+1}$  using \eqref{eq:updat_alpha} and \eqref{eq:update_p}. \\ 
\quad 9: \quad \quad Update $\hat{\sigma}^{m+1}_{0,i+1}$ and $\hat{\eta}^{m+1}_{1,i+1}$ using \eqref{eq:update_sigma} and \eqref{eq:update_eta}.   \\  
\quad 10:\quad \quad Construct $\hat{\vec{R}}^{m+1}_{i+1}$ with $\hat{\vec{p}}^{m+1}_{i+1}$, $\hat{\sigma}^{m+1}_{0,i+1}$ and $\hat{\eta}^{m+1}_{1,i+1}$. \\
\quad 11:\quad \quad Compute $\vec{g}^{m+1}_{i+1}$ with $\hat{\bm{\alpha}}^{m+1}_{i+1}$ and $\hat{\eta}_{1,i+1}^{m+1}$.\\
\quad 12:\quad \quad  $t_{i+1}=\frac{1+\sqrt{1+4t_{i}^2}}{2}$ \\
\quad 13:\quad \quad $\tilde{\vec{g}}^{m+1}_{i+1}= \vec{g}^{m+1}_i + \left( \frac{t_{i}-1}{t_{i+1}} \right) \vec{g}^{m+1}_{i+1}$   \\
\quad 14:\quad \quad  $i=i+1$ \\
\quad 15:\quad \quad  \textbf{until} practical convergence \\
\quad 16: \quad  $m=m+1$ \\
\quad 17: \quad  \textbf{until} practical convergence \\
\textbf{Output:} $\hat{\bm{\alpha}}$, $\hat{\sigma}_0$ and  $\hat{\eta}_1$. \\\\
\bottomrule
\end{tabular}
\end{table}
\end{small}
\subsection{Cyclically Refine the mLIKES Parameter Estimates} 
We first normalize the  mLIKES angle-Doppler estimate $\hat{\bm{\alpha}}$ by $\hat{\eta}_1$, i.e., we obtain $\hat{\bm{\alpha}}\hat{\eta}_1$. Denote $\bm{\chi}_k=\begin{bmatrix} \theta_k, \omega_k, \beta_{{\rm R},k}, \beta_{{\rm I},k} \end{bmatrix}$ as the parameters of the $k$-th strongest target. {The estimate of $\bm{\chi}_k$, i.e.,  $\hat{\bm{\chi}}_k$, can be determined  from the $k$-th maximum peak of the normalized mLIKES angle-Doppler image. We use the Bayesian information criterion \cite{stoica2004model} (BIC)  to determine the number of targets $K$. The cost function  of BIC for the mixed-ADC output is given by:
\begin{align}
    {\rm mBIC}(\breve{K})= -2 \ln{L\left(\lbrace \hat{\bm{\chi}}_k\rbrace_{k=1}^{\breve{K}},\eta_1\right)} + (6\breve{K}+1)\ln{(M_{\rm r}M_{\rm t}N)}. \label{eq:mbic}
\end{align} 
The estimate $\hat{K}$ of  the number  of targets is determined as the integer that minimizes the mBIC cost function with respect to the assumed number of targets $\breve{K}$.  Let $\mathcal{P}=\lbrace \hat{\bm{\chi}}_1,\dots, \hat{\bm{\chi}}_{\hat{K}}\rbrace$. } The angle-Doppler  parameter set $\mathcal{P}$ can be considered as the coarse estimates, which can be refined by Algorithm 2, where $\hat{\bm{\chi}}_k =\arg \min_{\bm{\chi}_k} -\ln{L\left( \bm{\chi}_k | \lbrace \hat{\bm{\chi}_j} \rbrace _{j=1,j\ne k}^{\hat{K}}, \hat{\eta} \right)}$, with $-\ln{L\left( \bm{\chi}_k | \lbrace \hat{\bm{\chi}_j} \rbrace _{j=1,j\ne k}^{\hat{K}},  \hat{\eta} \right)}$ given in Equation \eqref{eq:mle} with all but the $k$-th target parameters replaced with their estimates. This represents the parameter refinement of the $k$-th target while fixing the parameters of the other targets, e.g., $\lbrace \hat{\bm{\chi}_j} \rbrace _{j=1,j\ne k}^{\hat{K}}$. The noise parameter estimate $\eta$ is always updated along with  $\bm{\chi}_1$.  This refinement can be performed by using the interior-point based bounded optimization method (e.g., “fmincon” of MATLAB) over the angular interval $\left[ \hat{\theta}_k- \frac{90^{\circ}}{K_\theta}, \hat{\theta}_k + \frac{90^{\circ}}{K_\theta} \right]$ and the Doppler interval $\left[ \hat{\omega}_k- \frac{\pi}{K_\omega}, \hat{\omega}_k + \frac{\pi}{K_\omega} \right]$ to find the estimate of $\bm{\chi}_k$  that minimizes  $-\ln{L\left( \bm{\chi}_k | \lbrace \hat{\bm{\chi}_j} \rbrace _{j=1,j\ne k}^{\hat{K}}, \eta \right)}$. Note in passing that the idea of this cyclic refinement operation is similar to  the last step of the well-known RELAX algorithm \cite{li1996efficient,li1997angle}, and hence we refer to the two-step estimator that combines mLIKES and a variation of the last step of RELAX as mLIKES$\&$RELAX.   

{\subsection{Complexity Analysis} 
For the proposed mLIKES algorithm, the construction of $\hat{\vec{R}}_{m+1}^{i+1}$ and the computation of $(\hat{\vec{R}}_{m+1}^{i+1})^{-1}$ have complexities of $\mathcal{O}(N^2M_{\rm r}^2K_{\theta}K_{\omega})$ and  $\mathcal{O}(N^3M_{\rm r}^3)$, respectively.  The update of $\lbrace w_k\rbrace_{k=1}^{K_{\theta} K_{\omega}}$ in \eqref{eq:computew} has a complexity of $\mathcal{O}(N^2M_{\rm r}^2K_{\theta}K_{\omega})$. The update of $\bm{\alpha}$ in \eqref{eq:updat_alpha} requires a complexity of $\mathcal{O}(NM_{\rm r}K_{\theta}K_{\omega}+N^2M_{\rm r}^2)$  by first computing $\left(\hat{\vec{R}}^{m+1}_i\right)^{-1}\tilde{\vec{y}}^{m+1}_{i}$. The 
computational burden for updating $\vec{p}$, $\sigma_0$ and $\sigma_1$  is marginal and can be neglected.  Hence, the total computational complexity of mLIKES is on the order of $\mathcal{O}\left(I_1(N^2M_{\rm r}^2K_{\theta}K_{\omega}+N^3M_r^3)\right)$, where $I_1$ denotes the iteration number. It is worth noting that the proposed mLIKES algorithm for the mixed-ADC systems  has computational complexities similar to those of  the conventional LIKES algorithm for high-precision systems.  The following RELAX-based fine searches can be performed using the interior-point based bounded optimization method (e.g., “fmincon” of
MATLAB), where  the computational cost is proportional to $\mathcal{O}(\tilde{N}^{3.5})$, where $\tilde{N}$ is the number of design variables in $\bm{\chi}_k$. Assume that it requires $I_2$ iterations for RELAX-based algorithm to converge.  
 The total computational complexity of mLIKES$\&$RELAX is  on the order of  $\mathcal{O}\left(I_1(N^2M_{\rm r}^2K_{\theta}K_{\omega}+N^3M_r^3)+I_2\tilde{N}^{3.5}\right)$.}

\begin{table}[htb]
\renewcommand\arraystretch{1.0}
\centering
\begin{tabular}{l}
\toprule
\textbf{Algorithm 2:} Cyclically refine the parameters \\
\midrule
\textbf{Input:} $\mathcal{P}=\lbrace \hat{\bm{\chi}}_1, \dots, \hat{\bm{\chi}}_{\hat{K}} \rbrace$:  Coarse estimates of the  \\ 
\qquad \qquad  parameters obtained from mLIKES \\
\quad \quad \quad ${\hat{K}}$: Estimated target number\\
\textbf{Procedure:}\\
\quad 1: \textbf{repeat}\\
\quad 2: \quad $\hat{\bm{\chi}}_1,\hat{\eta}=\arg \min_{\bm{\chi}_1,\eta} -\ln{L\left( \bm{\chi}_1,\eta | \lbrace \hat{\bm{\chi}_j} \rbrace _{j=2}^{\hat{K}} \right)}$ \\
\quad 3: \quad \textbf{for} $k=2,\dots,\hat{K}$ \\
\quad 4: \quad \quad \quad $\hat{\bm{\chi}}_k=\arg \min_{\bm{\chi}_k} -\ln{L\left( \bm{\chi}_k | \lbrace \hat{\bm{\chi}_j} \rbrace _{j=1,j\ne k}^{\hat{K}},\hat{\eta} \right)}$ \\
\quad 5: \quad \textbf{end for}\\
\quad 6: \textbf{until} practical convergence \\
\textbf{Output:}$\begin{bmatrix}\hat{\theta}_1, \hat{\omega}_k,\hat{b}_k=\hat{\beta}_k/\hat{\eta}\end{bmatrix}$ for $k=1,\dots,\hat{K}$ and $\hat{\eta}$. \\\\
\bottomrule
\end{tabular}
\end{table}
\vspace{-0.2cm}
\section{Numerical examples}
In this section, we present several numerical examples to demonstrate the performance of the mixed-ADC based architecture and the proposed algorithms for PMCW MIMO radar angle-Doppler Imaging. The PMCW MIMO radar under consideration is equipped with $M_{\rm t}=10$ transmit antennas spaced at $d_{\rm t}=5\lambda$ and $M_{\rm r}=10$ receive antennas spaced at $d_{\rm r}=\lambda / 2$. With the filled receive  array and  sparse transmit array, we can effectively create a filled virtual array with 100 antennas, see \eqref{eq:vecmodel}.  The slow-time sample number $N=64$ PRI's are adopted to extract the Doppler information. Random binary sequences (i.e., $C(m,n)\in \lbrace 1,-1\rbrace$ with equal probabilities) are used  as the slow-time codes.  All examples were run on a PC with Intel(R) Core(TM) i7-6700 CPU @ 3.40GHz  and 64.0 GB RAM.

\subsection{Cram{\'e}r-Rao Bound for Mixed-ADC Based Receiver} \label{sec:secA}
For the mixed-ADC based architecture, we consider the following three situations\footnote{For a fixed $L$, our extensive numerical simulations demonstrate that the angle-Doppler estimation performance is not sensitive to the distribution of high-precision ADC’s in a MIMO system.}: 
\begin{enumerate}
\item[1.] $\delta_1=1$ and $\lbrace \delta_m=0\rbrace_{m=2}^{10}$, which means that the first receive antenna is equipped with a pair of high precision ADC's (i.e., for in-phase and quadrature (I/Q) branches) and all other receive antennas are  equipped with one-bit (1b) ADC's;
\item[2.] $\delta_1=1$, $\delta_2=1$ and $\lbrace \delta_m=0\rbrace_{m=3}^{10}$, which indicates that the first and  second of the receive antennas are equipped with high-precision  ADC's and the other 8 with one-bit ADC's;
\item[3.]$\lbrace \delta_m=1 \rbrace_{m=1}^5$ and $\lbrace \delta_m=0\rbrace_{m=6}^{10}$.
\end{enumerate}
\par We consider a case of $K=2$  targets. First, we assume that, $\theta_1=22.5^{\circ}$, $\theta_2=25.3125^{\circ}$, $\omega_1=1.3$, $\omega_2=1.4$, $b_1=1e^{j\pi/4}$ and $b_2=e^{j\pi/4}/r$. {Therefore $r$ represents the  amplitude ratio between the strong and weak targets:
\begin{equation}
    r=\frac{|b_1|}{|b_2|}.
\end{equation}} 
We vary $r$ from 1 to 1000. We also vary the noise variance to maintain the same ${\rm SNR}_2$ = 10 dB for the weak target, for all values of $r$, where: 
\begin{equation}
    {\rm SNR}_2=10\log_{10} \frac{1}{r^2\sigma^2} {\rm dB}. 
\end{equation}

For the one-bit ADC system and the one-bit part of the mixed-ADC based architecture, the PRI-varying (slow-time varying) threshold has the real and imaginary parts selected randomly and equally likely from a predefined eight-element set $\lbrace -h_{\rm max}, -h_{\rm max}+\Delta, \dots, h_{\rm max}-\Delta, h_{\rm max} \rbrace$ with $h_{\rm max}=\sqrt{p_{\rm out}}$  and $\Delta=h_{\rm max}/7$, where $\sqrt{p_{\rm out}}$ is the average received signal power at the I/Q channels.  
\par
\begin{figure}[htb]
\centering
\subfigure[Root CRB($\omega_1$)]{
\label{fig:CRB_w1} 
\begin{minipage}[t]{0.48\linewidth}
\centering
\centerline{\epsfig{figure=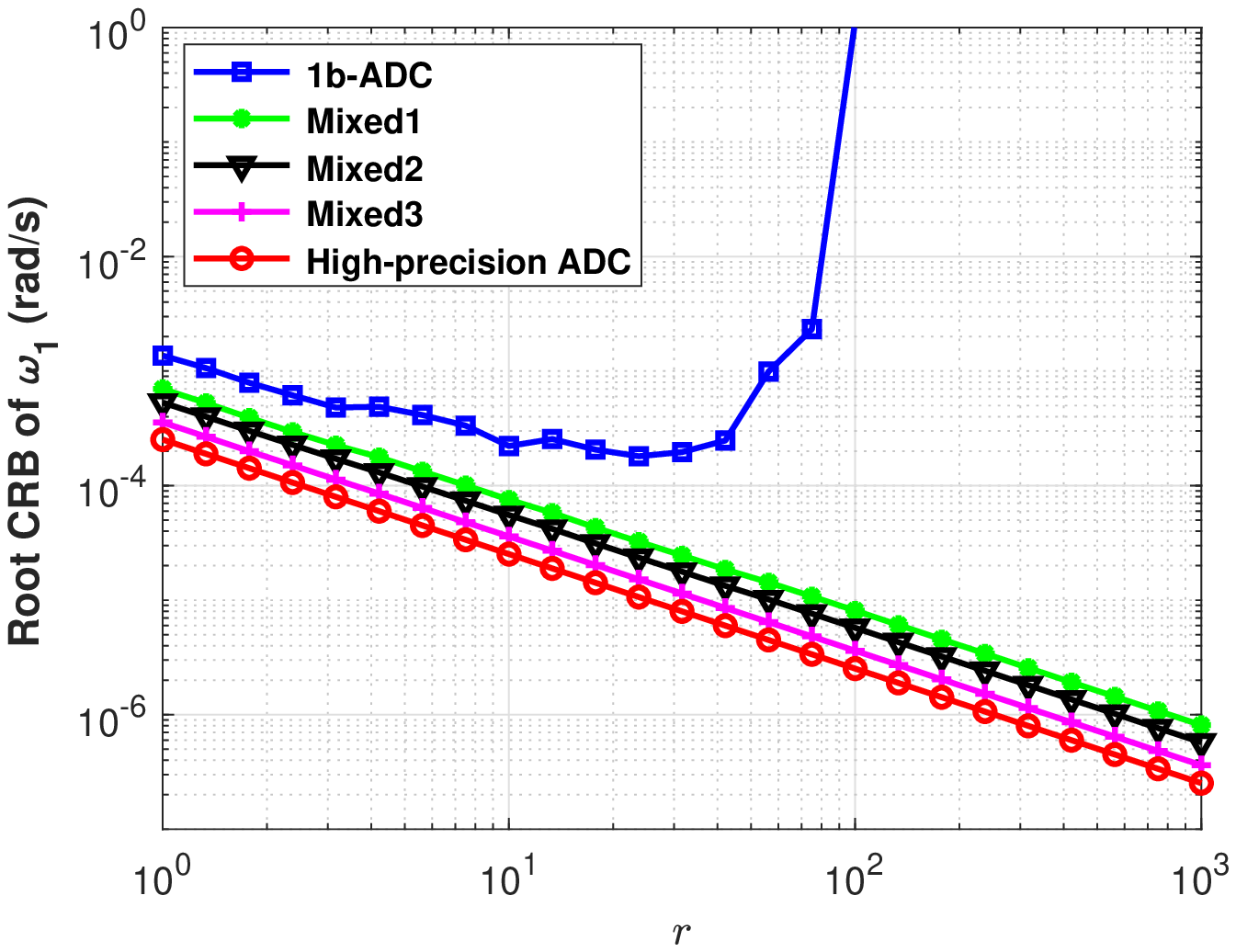,width=4.6cm}}
\end{minipage}} 
\subfigure[Root CRB($\theta_1$)]{
\label{fig:CRB_theta1}
\begin{minipage}[t]{0.48\linewidth}
\centering
\centerline{\epsfig{figure=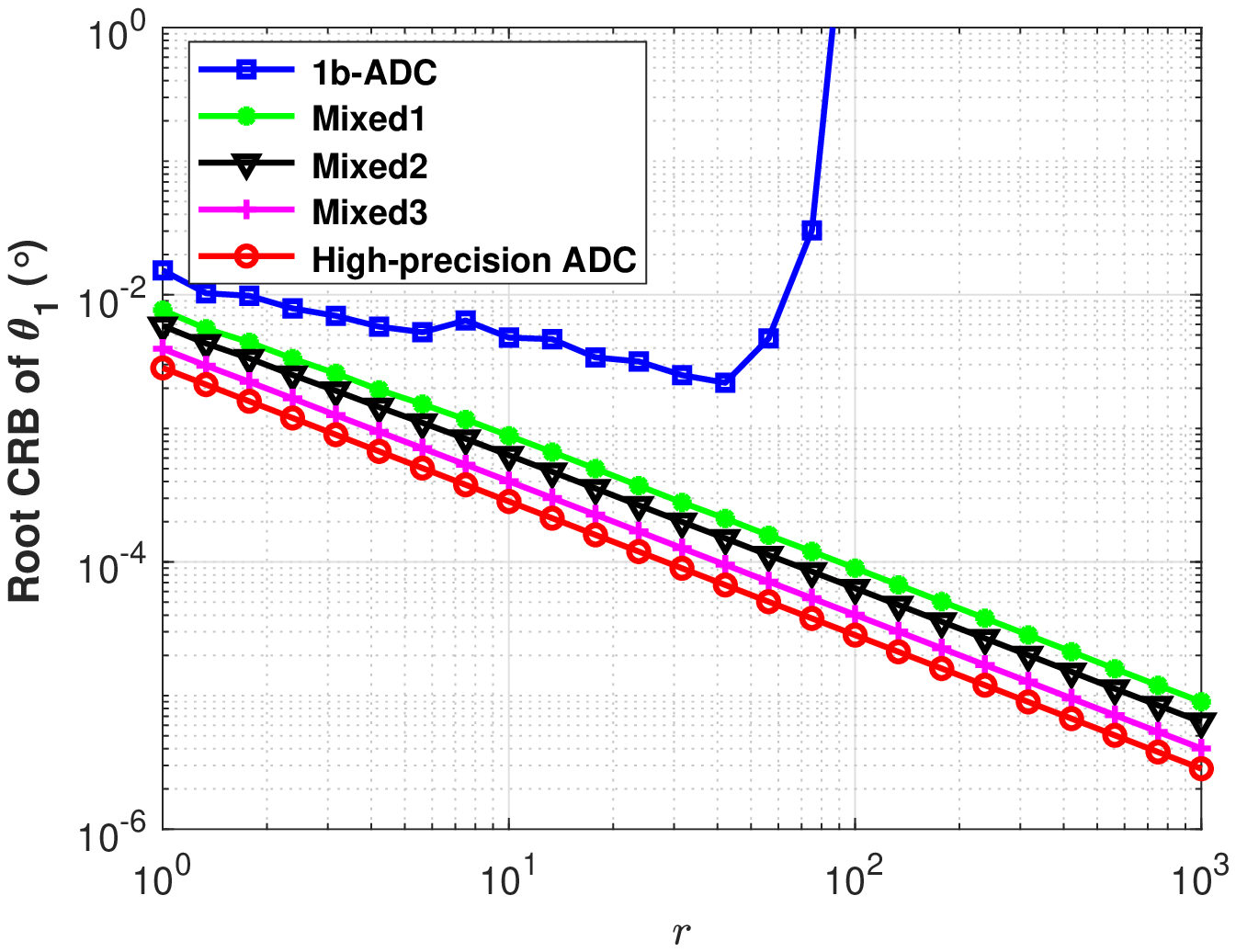,width=4.6cm}}
\end{minipage}}
\caption{Root CRB versus $r$ for (a) $\omega_1$ and (b) $\theta_1$.}
\label{fig:crb1}
\end{figure}

\begin{figure}[htb]
\centering
\subfigure[Root CRB($\omega_2$)]{
\label{fig:CRB_w2}
\begin{minipage}[t]{0.48\linewidth}
\centering
\centerline{\epsfig{figure=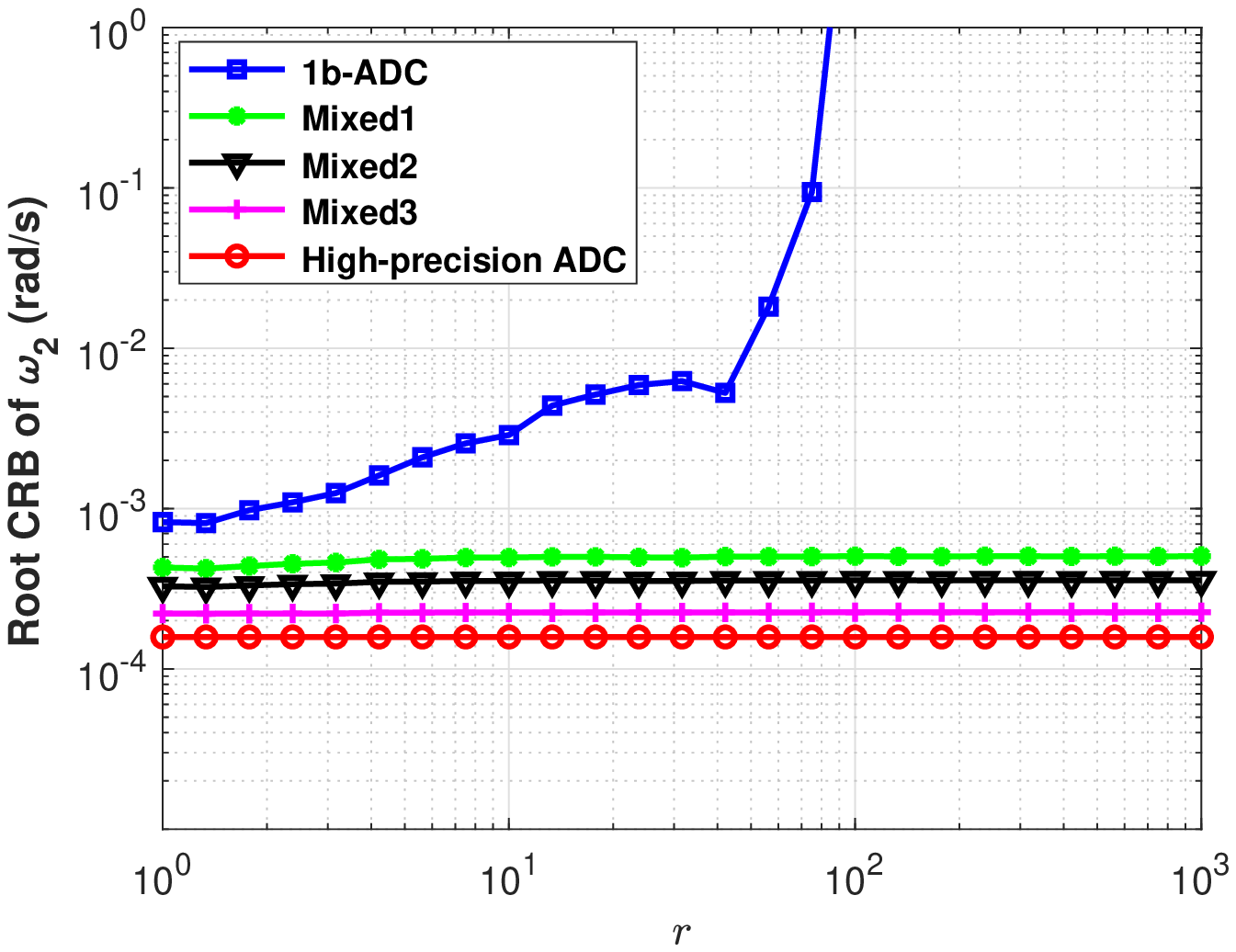,width=4.6cm}}
\end{minipage}} 
\subfigure[Root CRB($\theta_2$)]{
\label{fig:CRB_theta2}
\begin{minipage}[t]{0.48\linewidth}
\centering
\centerline{\epsfig{figure=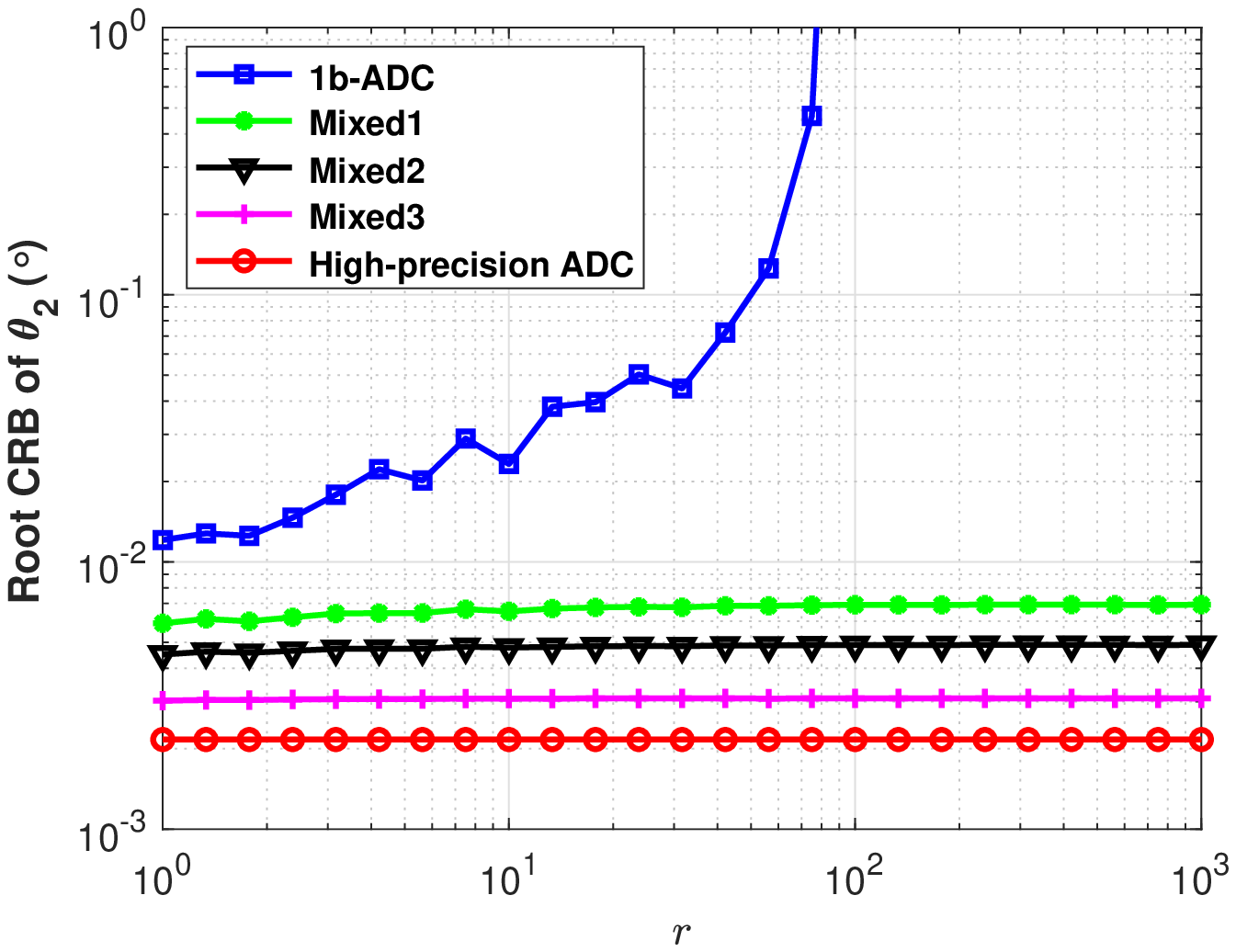,width=4.6cm}}
\end{minipage}}
\caption{Root CRB versus $r$ for (a) $\omega_2$  and (b) $\theta_2$.}
\label{fig:crb2}
\end{figure}

 Figs. \ref{fig:crb1} and \ref{fig:crb2} show the root CRBs (RCRBs) for $\bm{\theta}$ and $\bm{\omega}$ (see the explanations in the figure captions). 
Note from Fig. \ref{fig:crb1} that the high-precision root CRBs for $\omega_1$ and $\theta_1$, decrease with $r$ because $\sigma^2$ decreases as $r$ increases and hence ${\rm SNR}_1$ increases with $r$.  For the weak target, the high-precision root CRBs in Fig. \ref{fig:crb2} for $\omega_2$ and $\theta_2$ are constant as $r$ varies because ${\rm SNR}_2$ is the same for all values of $r$. Note that the high-precision sampling receiver does not suffer from  dynamic range problems even when $r$ is very large. For $r=1000$ for example, the first target is 60 dB stronger than the second one.  

\par The 1b-based CRB in Figs. \ref{fig:crb1} and \ref{fig:crb2} for $\omega_1$, $\omega_2$, $\theta_1$ and $\theta_2$ are the largest of all the cases considered. This is the price paid for the low-cost of the binary quantizers, and  the increases of the root CRB's  are significant and can reach unacceptable levels. This is the so-called dynamic range problem of coarse quantizations. Compared with the one-bit system, mixed-ADC based architectures (Situations 1, 2 and 3  are denoted as ``Mixed1'', ``Mixed2'' and ``Mixed3'', respectively) with just one pair of high precision ADC's can attain significant performance improvements, especially for large values of $r$ (e.g., $r\geq 10$). Most notably, the root CRB's of $w_2$ and $\theta_2$ remain almost constant  as $r$ increases, and hence the mixed-ADC based architecture can be a viable solution to solve the dynamic range problems of using coarse quantizers to reduce cost and power consumption of PMCW MIMO radar. We note that  with just one pair of high-precision ADC's, the mixed-ADC based system can attain a dynamic range as high as 60 dB without suffering from significant accuracy problems. {Therefore the mixed-ADC architecture with one pair of high-precision ADC's and a large number of one-bit ADC's is a low-cost solution for PMCW radar. }
\subsection{Angle-Doppler Imaging}
Finally, we present several numerical examples to demonstrate the angle-Doppler imaging performance of the proposed algorithm for the mixed-ADC based PMCW MIMO radar system.
\begin{figure}[htb]
\centering
\subfigure[RMSE of $\theta_2$]{
\label{fig:w2}
\begin{minipage}[t]{0.48\linewidth}
\centering
\centerline{\epsfig{figure=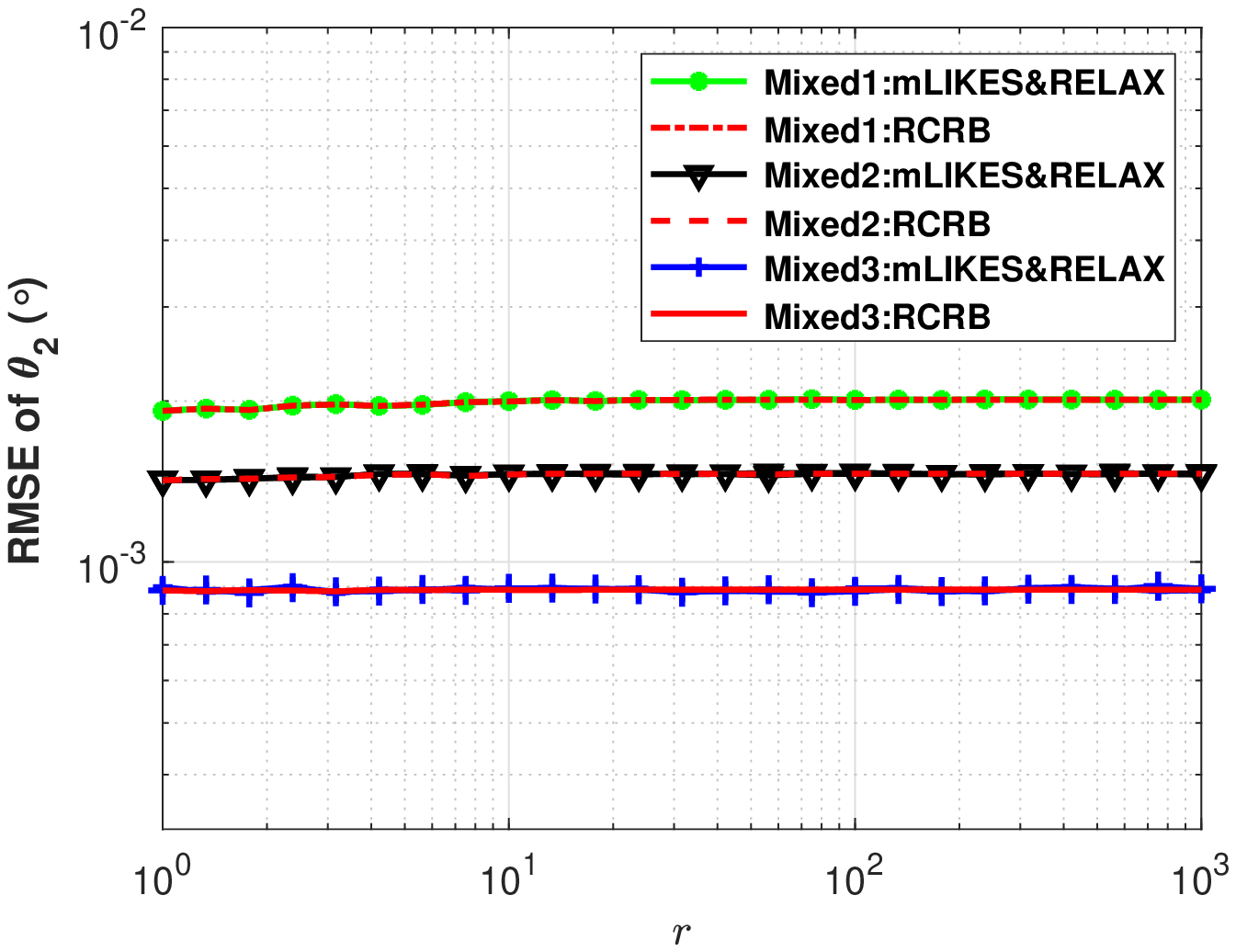,width=4.6cm}}
\end{minipage}}
\subfigure[RMSE of $\omega_2$]{
\label{fig:theta2}
\begin{minipage}[t]{0.48\linewidth}
\centering
\centerline{\epsfig{figure=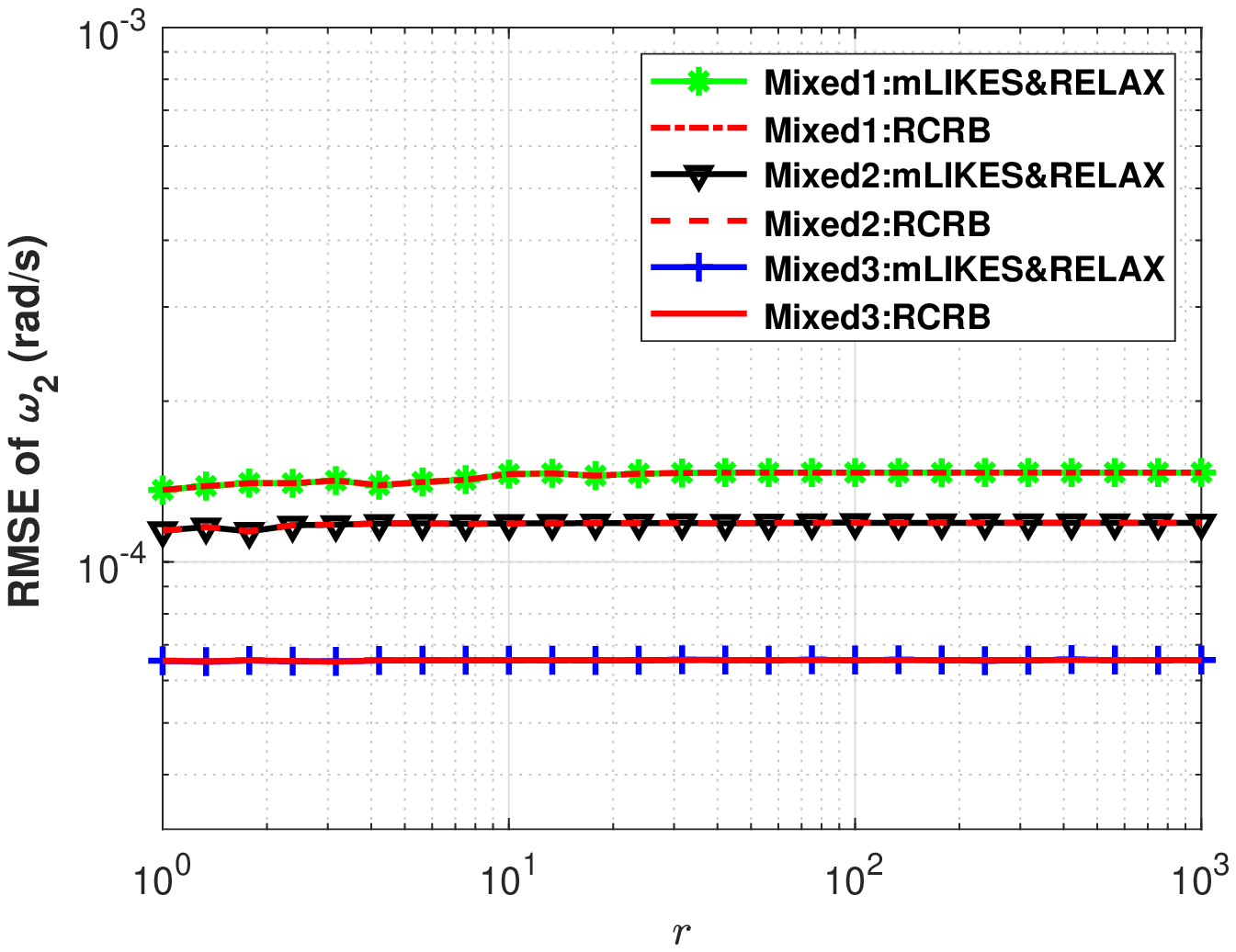,width=4.6cm}}
\end{minipage}} 
\subfigure[RMSE of $b_{2,{\rm R}}$]{
\label{fig:b2R}
\begin{minipage}[t]{0.48\linewidth}
\centering
\centerline{\epsfig{figure=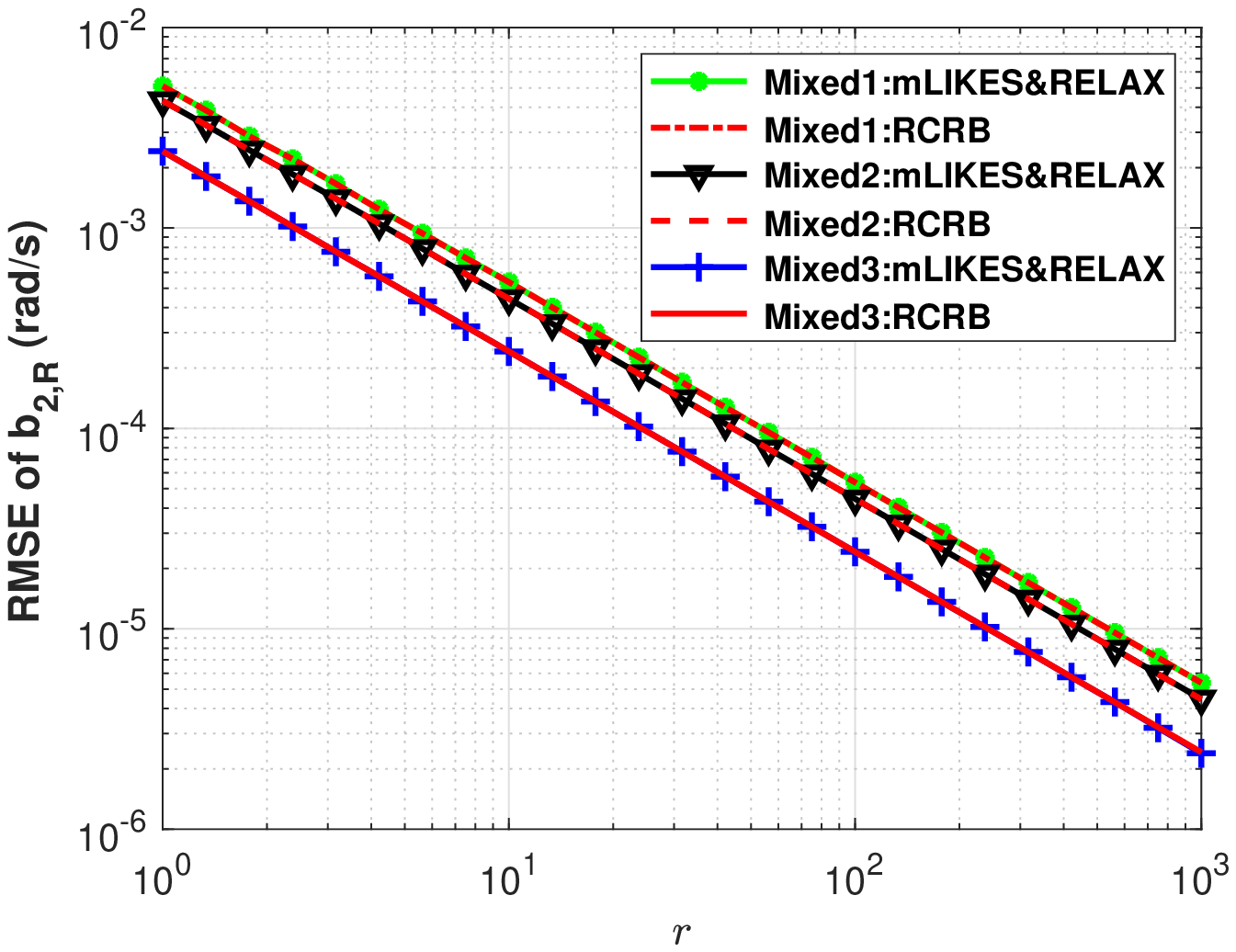,width=4.6cm}}
\end{minipage}}
\subfigure[RMSE of $b_{2,{\rm I}}$]{
\label{fig:b2I}
\begin{minipage}[t]{0.48\linewidth}
\centering
\centerline{\epsfig{figure=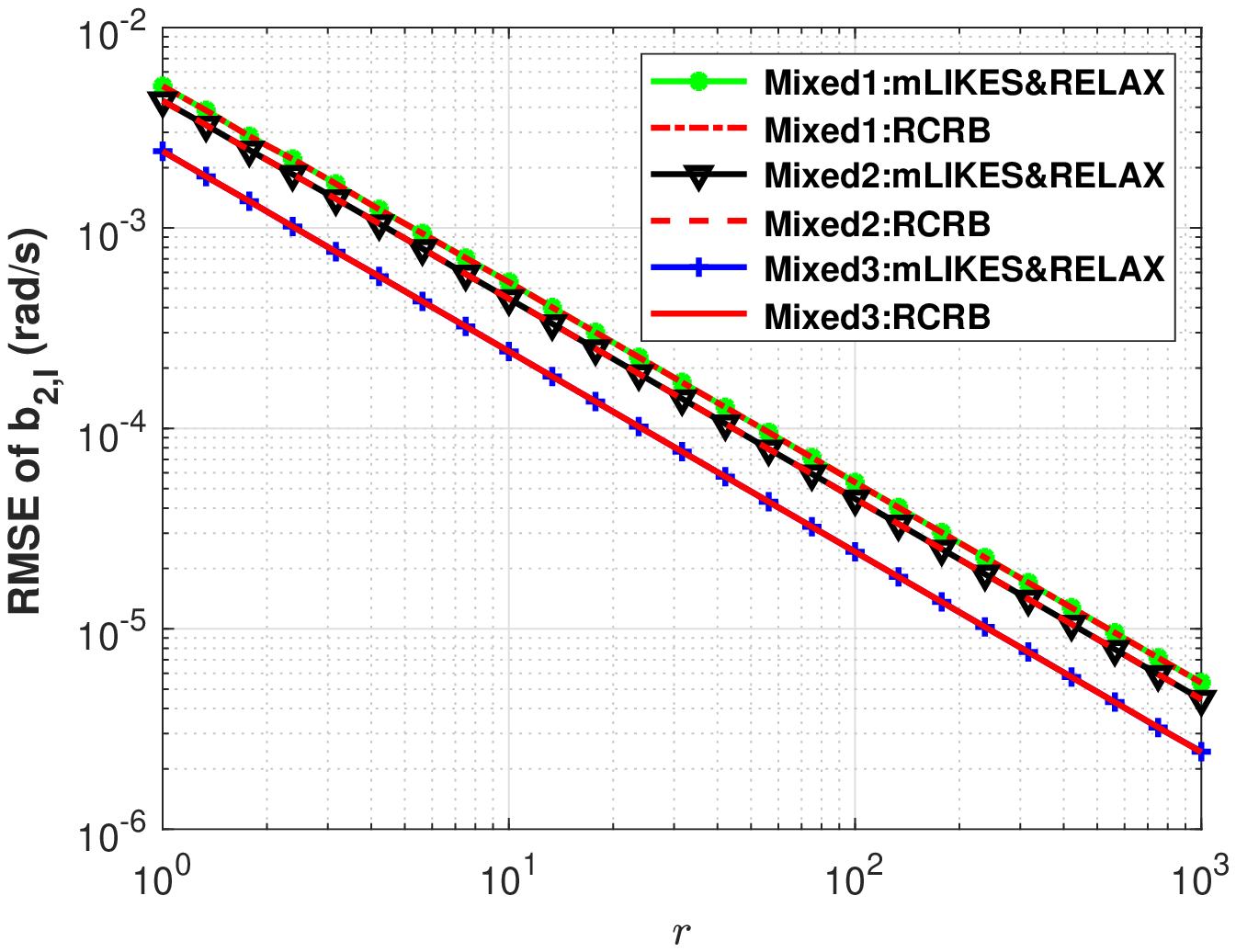,width=4.6cm}}
\end{minipage}}
\caption{Comparison between RCRBs and RMSEs of the parameter estimates of the second target as a function of $r$. } 
\label{fig:rmse}
\end{figure}
\subsubsection{Implementation details}
The number of grid points in the angular and Doppler domains are set to $K_{\theta}$ =128 and $K_{\omega}=256$, respectively. In our simulations, the practical convergence of mLIKES is considered achieved when the relative change of $\vec{p}$ between two consecutive iterations is below a small threshold.  For the outer MM iterations, we use the threshold $\varepsilon_1= 10^{-4}$ and we terminate the iterations when a maximum iteration number $I_1$ = 50 is reached. For the inner MM iterations in mLIKES, we also set the threshold to $\varepsilon_2=10^{-4}$ and we terminate the iterations when a maximum iteration number $I_2=10$ is reached. For the RELAX-based cyclic refinement, we stop the iterations when the relative  change  of the negative log-likelihood function between two consecutive iterations is below $10^{-6}$ or a maximum iteration number reaching 50. 

\subsubsection{Example 1} 
Figs. \ref{fig:w2}-\ref{fig:b2I} show the root mean-squared errors (RMSEs) of the second target parameters obtained with 500 Monte-Carlo trials as a function of $r$. (The RMSEs of the first target parameter estimates obtained by using mLIKES$\&$RELAX  are close to the RCRB, and the plots are not shown herein.)  It is observed from Fig. \ref{fig:rmse} that the RMSEs of the estimates obtained by using mLIKES$\&$RELAX  can approach the CRB for all  values of $r$ considered, i.e., for a dynamic range up to 60 dB.   

\subsubsection{Example 2} 
\begin{figure*}[htb]
\centering
\subfigure[MF]{
\label{fig:pmcw_das}
\begin{minipage}[t]{0.43\linewidth}
\centering
\centerline{\epsfig{figure=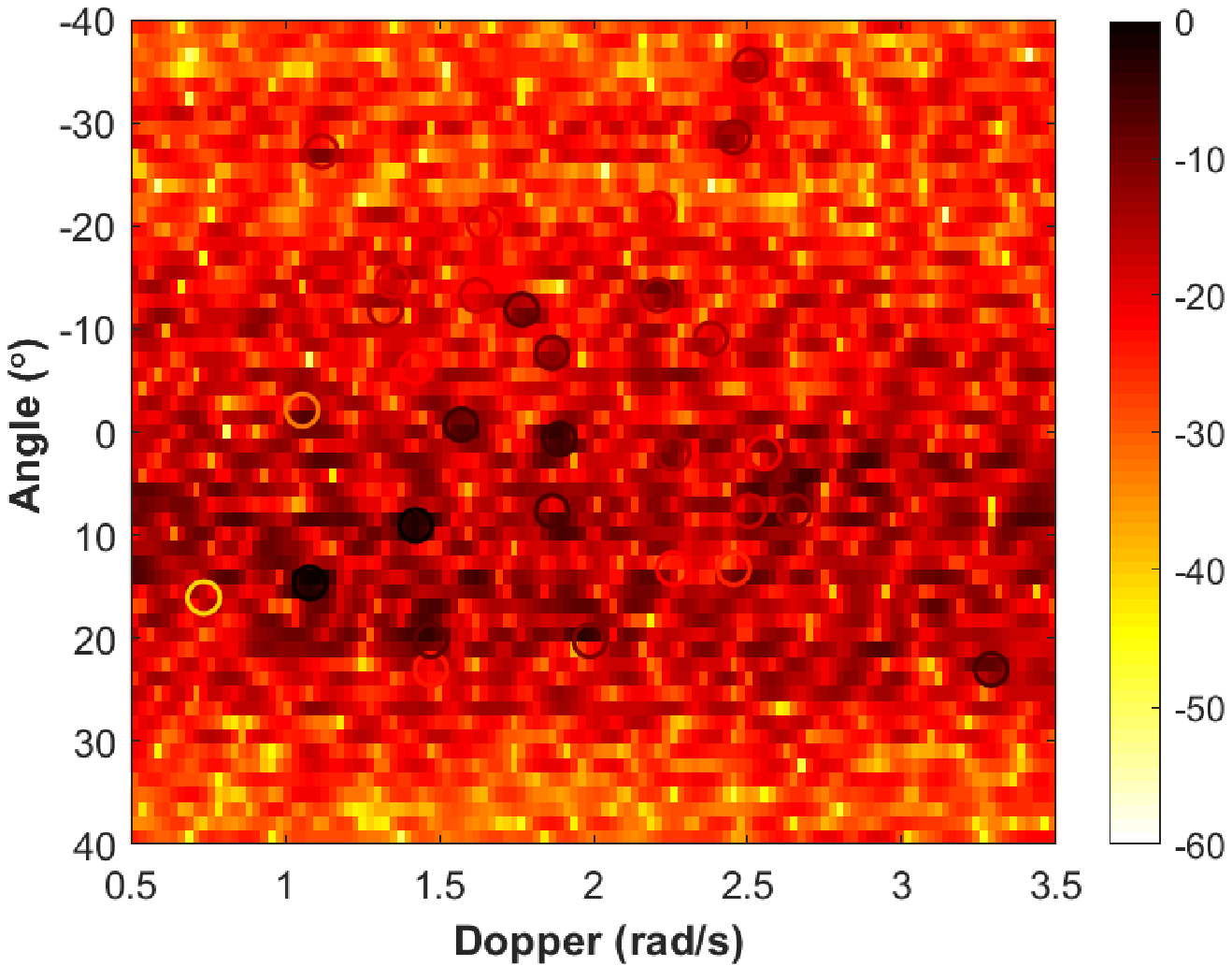,width=7.2cm}}
\end{minipage}}
\subfigure[LIKES]{
\label{fig:pmcw_likes}
\begin{minipage}[t]{0.43\linewidth}
\centering
\centerline{\epsfig{figure=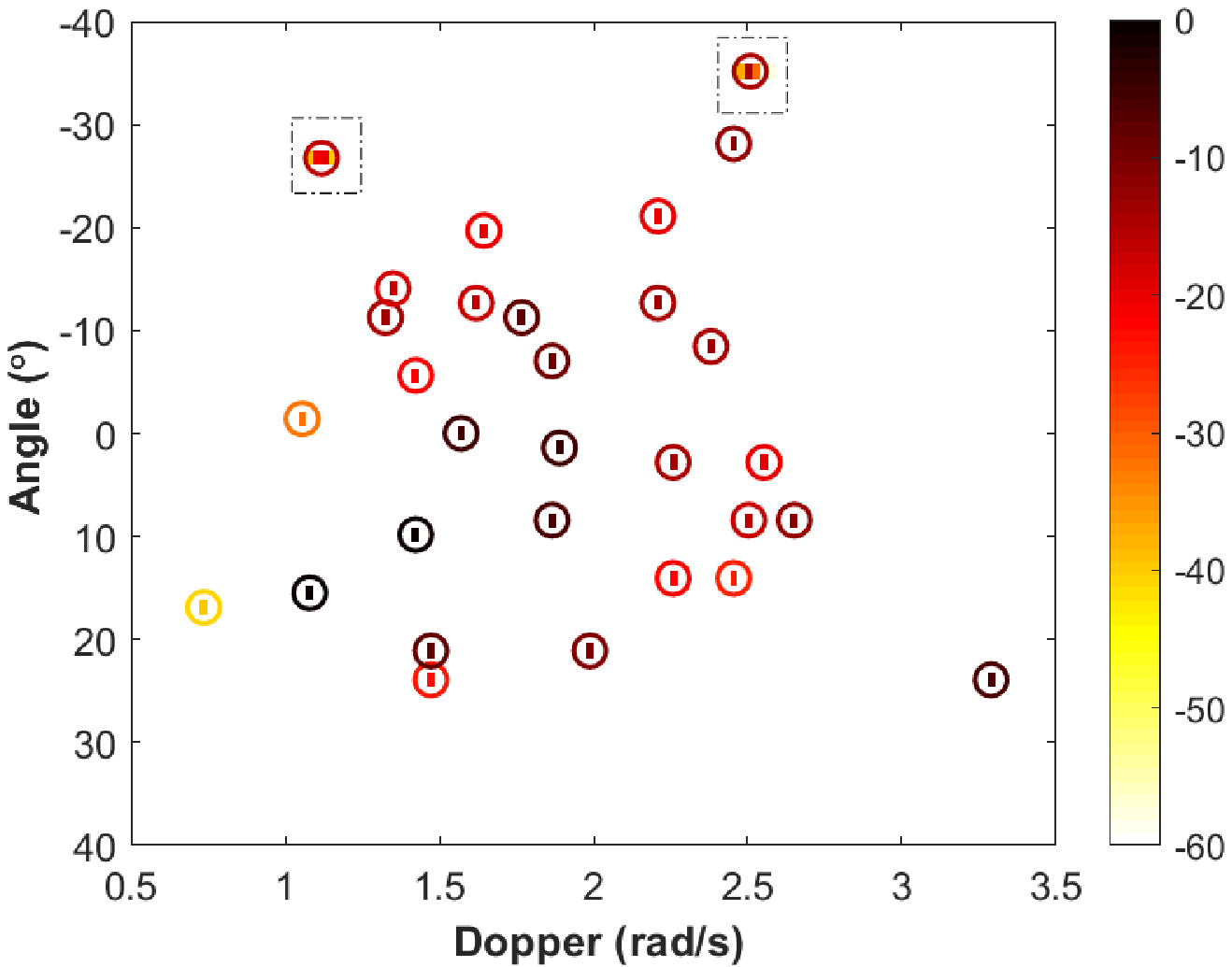,width=7.2cm}}
\end{minipage}}
\subfigure[1bLIKES]{
\label{fig:fmcw_1blikes}
\begin{minipage}[t]{0.43\linewidth}
\centering
\centerline{\epsfig{figure=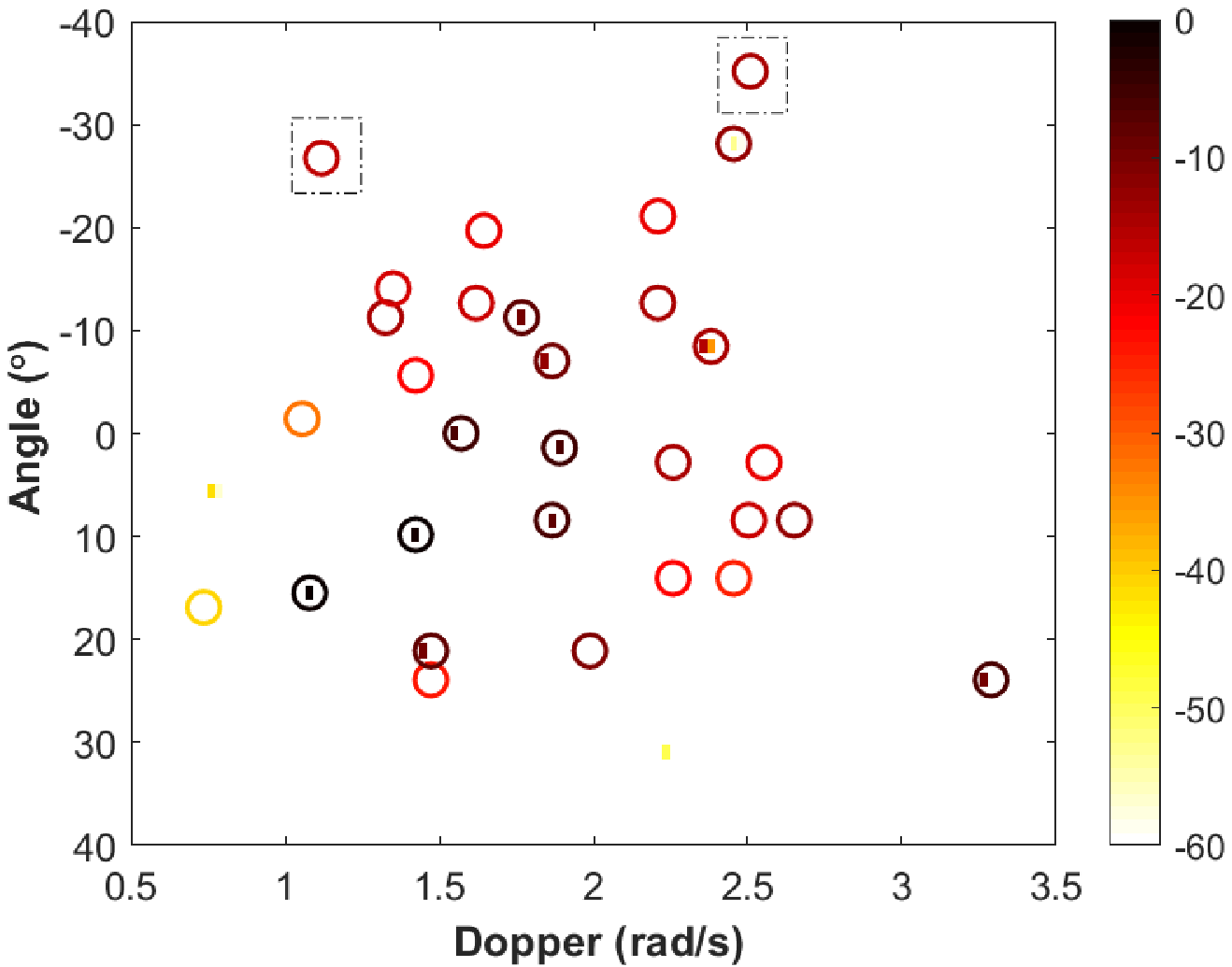,width=7.2cm}}
\end{minipage}}
\subfigure[mLIKES]{
\label{fig:pmcw_mlikes}
\begin{minipage}[t]{0.43\linewidth}
\centering
\centerline{\epsfig{figure=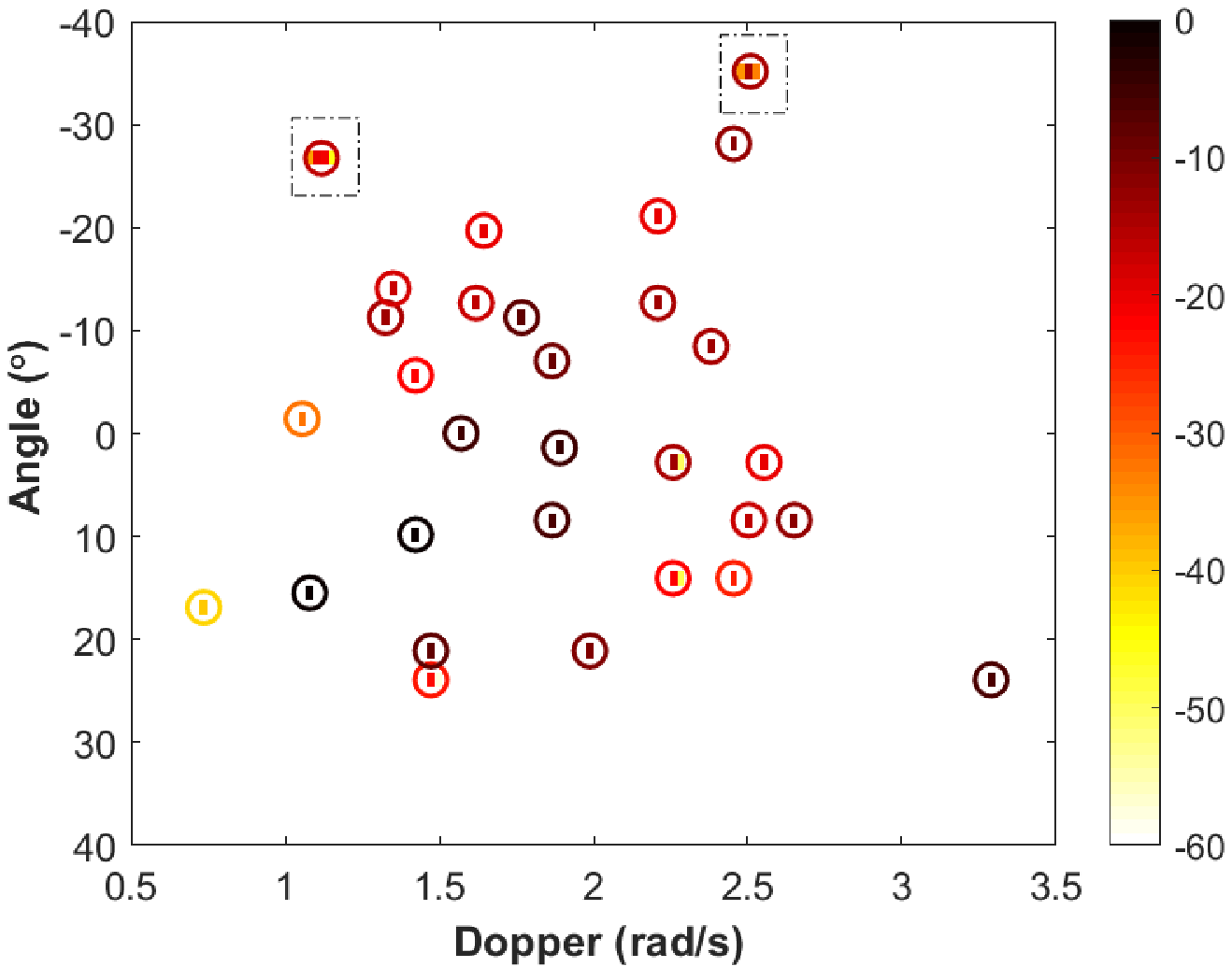,width=7.2cm}}
\end{minipage}}
\centering
\subfigure[mLIKES + cyclic refinement]{
\label{fig:pmcw_mrelax}
\begin{minipage}[t]{0.43\linewidth}
\centering
\centerline{\epsfig{figure=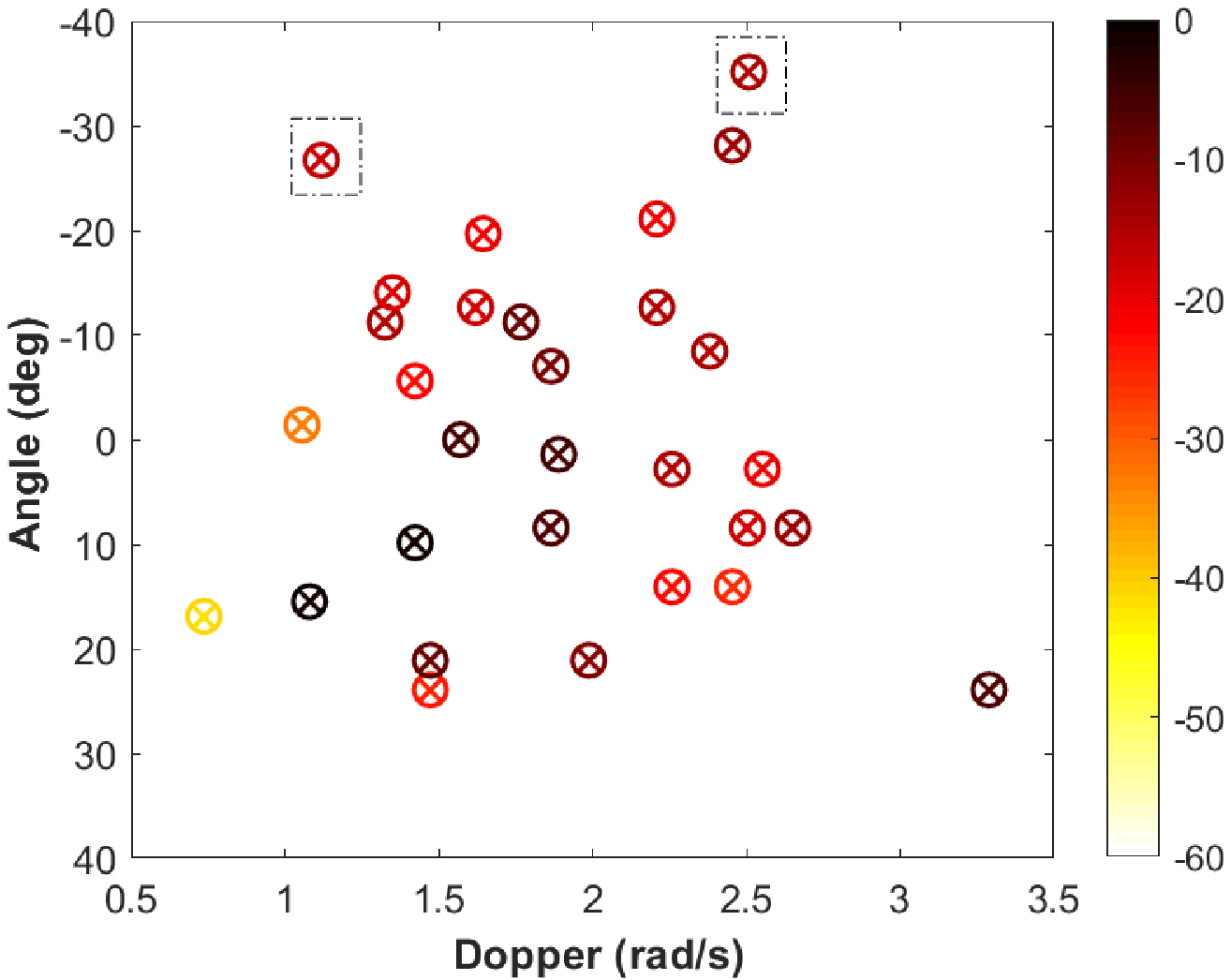,width=7.2cm}}
\end{minipage}}
\caption{Ange-Doppler images of mixed-ADC based PMCW MIMO radar, obtained by applying (a) matched filter, (b) LIKES to  the high-precision data, (c) 1bLIKES to the one-bit data and (d) mLIKES and (e) mLIKES$\&$RELAX to the Mixed1 outputs.  ``$\bigcirc$'' symbols indicate  the locations of the true targets (color-coded according to power, in dB). All power levels are in dB. The dash-dot rectangles in (b)-(e) indicate the off-grid targets. ``X'' symbols in Fig. \ref{fig:pmcw_mrelax} represent  targets estimated by  mLIKES$\&$RELAX.} 
\label{fig:pmcw_imaging}
\end{figure*}

We now consider 30 moving targets  with their angle-Doppler locations and powers indicated by the color-coded ``$\bigcirc$'', as shown in Fig. \ref{fig:pmcw_imaging}. Note that there are two off-grid targets marked with dash-dot rectangles. The amplitudes of the targets $\lbrace |b_k| \rbrace_{k=1}^{30}$ are selected randomly between 0.01 and 1, resulting in a dynamic range of 40 dB. The additive noise is assumed to be circularly symmetric i.i.d. complex-valued  white Gaussian noise, with mean zero and a variance resulting in the minimum target SNR of 10 dB. Similar to Section \ref{sec:secA}, the 8-level PRI-varying thresholds are considered for the one-bit system and the one-bit part of the mixed-ADC based architecture. 

The angle-Doppler images obtained by using the matched filter  with the high-precision data is shown in Fig. \ref{fig:pmcw_das}. Since the waveform orthogonality is not perfectly achieved in the Doppler domain, the slow-time code residuals are dispersed into the entire angle-Doppler images as pseudo noise. As a result, most weak targets are masked by the slow-time code residuals. However, we can observe from Fig. \ref{fig:pmcw_likes} that the LIKES algorithm can  estimate the strong targets accurately and the weak targets reasonably well. In comparison, 1bLIKES misses quite a few weak targets, which is primarily due to the fact that the estimation performance offered by the one-bit quantizer degrades dramatically when the dynamic range of the signal components is high \cite{stoica2021cramer}. However, after introducing just one pair of high-precision ADC's, used at a single antenna output, into the one-bit ADC system, the dynamic range is drastically improved and mLIKES produces a satisfactory angle-Doppler image with all targets identified.  Note from Figs. \ref{fig:pmcw_likes} and \ref{fig:pmcw_mlikes} that the two off-gird targets are slightly smeared.  Fig. \ref{fig:pmcw_mrelax} shows that the cyclic refinement operation improves upon the mLIKES  results and accurately determines the angle-Doppler locations of the two off-grid targets. {The computational times needed by these methods in the present example  are as follows: MF$-$0.02 seconds, LIKES$-$71 seconds, 1bLIKES$-$252 seconds, mLIKES$-$167 seconds, and mLIKES$\&$RELAX$-$187 seconds.}

\section{Conclusions}
We have considered a  mixed-ADC based architecture for  PMCW  MIMO radar systems. We have derived the CRB for the system  to characterize its  best achievable unbiased estimation performance of target parameters.   By making use of the MM technique, a computationally efficient estimator, referred to as mLIKES, has been introduced to obtain accurate angle-Doppler images. To further enhance the target parameter estimation performance, a RELAX-based approach is used to cyclically refine the  mLIKES results  to realize the ML estimation.  Numerical examples have been presented to illustrate that the mixed-ADC based architecture with just one pair of high-precision ADC's at a single antenna output allows us to significantly reduce the hardware cost and power consumption while still maintain a high dynamic range needed by the automotive radar for autonomous driving applications. We have also demonstrated that the proposed algorithms can be used to attain  good  angle-Doppler imaging performances. 

\appendices
\section{Cram{\'e}r-Rao bound for high-precision ADC's} \label{appendix:A}
Note from \cite{stoica2021cramer}\cite{li2007range} that the CRB formula for the data model in \eqref{eq:model} is given by:
\begin{equation}
    F(\varphi_i,\varphi_j) = \frac{2}{\sigma_0^2} \Re\left\{ {\rm tr} \left[ \frac{\partial (\vec{A}_{\rm r}\vec{B}\vec{V}^T)^H}{\partial \varphi_i}\frac{\partial (\vec{A}_{\rm r}\vec{B}\vec{V}^T)}{\partial \varphi_j} \right] \right\}. 
\end{equation}
The derivative of $\bm{\theta}$, $\bm{\omega}$, $\Re(\vec{b})$ and $\Im(\vec{b})$ with respect to $\vec{A}_{\rm r}\vec{B}\vec{V}^T$ gives:
\begin{align}
    \frac{\partial (\vec{A}_{\rm r}\vec{B}\vec{V}^T)}{\partial \theta_i} & =  \dot{\vec{A}}_{\rm r}\vec{e}_i\vec{e}_i^T\vec{B}\vec{V}^T + \vec{A}_{\rm r}\vec{B}\vec{e}_i\vec{e}_i^T\dot{\vec{V}}^T_{\vec{\theta}}, \nonumber \\
    \frac{\partial (\vec{A}_{\rm r}\vec{B}\vec{V}^T)}{\partial \omega_i} & = \vec{A}_{\rm r}\vec{B}\vec{e}_i\vec{e}_i^T\dot{\vec{V}^T}_{\omega}, \nonumber \\
    \frac{\partial (\vec{A}_{\rm r}\vec{B}\vec{V}^T)}{\partial b_{{\rm R},i}}&= \vec{A}_{\rm r}\vec{e}_i\vec{e}_i^T\vec{V}^T, \nonumber \\
    \frac{\partial (\vec{A}_{\rm r} \vec{B}\vec{V}^T)}{\partial b_{{\rm I},i}}&= j \vec{A}_{\rm r} \vec{e}_i\vec{e}_i^T\vec{V}^T,
\end{align}
where $\dot{\vec{A}}_{\rm r}$, $\dot{\vec{V}}_{\vec{\theta}}$ and $\dot{\vec{V}}_{\bm{\omega}}$ are defined in \eqref{eq:dAr}$-$\eqref{eq:dAf}, respectively. Then
\begin{align}
    F(\theta_i,\theta_j) = \frac{2}{\sigma_0^2} \Re \left\{ {\rm tr} \left[ ( \dot{\vec{A}}_{\rm r}\vec{e}_i\vec{e}_i^T\vec{B}\vec{V}^T + \vec{A}_{\rm r}\vec{B}\vec{e}_i\vec{e}_i^T\dot{\vec{V}}^T_{\vec{\theta}}  )^H \right. \right.   \nonumber \\ 
    \left. \left.    \times ( \dot{\vec{A}}_{\rm r}\vec{e}_j\vec{e}_j^T\vec{B}\vec{V}^T + \vec{A}_{\rm r}\vec{B}\vec{e}_j\vec{e}_j^T\dot{\vec{V}}^T_{\vec{\theta}} ) \right] \right\}. \label{eq:Ftheta}
\end{align}
Next note that
\begin{align}
    {\rm tr}&\left[ (\dot{\vec{A}}_{\rm r}\vec{e}_i\vec{e}_i^T\vec{B}\vec{V}^T)^H (\dot{\vec{A}}_{\rm r}\vec{e}_j\vec{e}_j^T\vec{B}\vec{V}^T) \right] \nonumber \\
    &= (\vec{e}_i^T\dot{\vec{A}}_{\rm r}\dot{\vec{A}}_{\rm r}\vec{e}_j) (\vec{e}_j^T \vec{B}^{*}\vec{V}^{*}\vec{V}^T\vec{B}\vec{e}_i) \nonumber \\
    &= (\dot{\vec{A}}_{\rm r}\dot{\vec{A}}_{\rm r})_{i,j} (\vec{B}^{*}\vec{V}^H\vec{V}\vec{B})_{i,j},
\end{align}
where we have used the fact that ${\rm tr}(\vec{A}\vec{B}\vec{C})={\rm tr}(\vec{B}\vec{C}\vec{A})$. The other three matrix product terms in \eqref{eq:Ftheta} have similar forms, and hence $\vec{F}_{11}$ has the form in \eqref{eq:F11}. Similar to the derivation of $\vec{F}_{11}$, the other sub-matrices of the FIM follow immediately and are given in \eqref{eq:F12}$-$\eqref{eq:F33}.

\small 
\bstctlcite{IEEEexample:BSTcontrol}
\bibliographystyle{IEEEtran} 
\normalsize
\bibliography{main}
\end{document}